\documentclass[prd,preprint,showpacs]{revtex4}

\usepackage{graphicx}
\usepackage{epstopdf}
\usepackage{latexsym}
\usepackage{amssymb}
\usepackage{color}

\usepackage[center]{subfigure}

\begin{document}

 \newcommand{\bq}{\begin{equation}}
 \newcommand{\eq}{\end{equation}}
 \newcommand{\bqn}{\begin{eqnarray}}
 \newcommand{\eqn}{\end{eqnarray}}
 \newcommand{\nb}{\nonumber}
 \newcommand{\lb}{\label}
\newcommand{\PRL}{Phys. Rev. Lett.}
\newcommand{\PL}{Phys. Lett.}
\newcommand{\PR}{Phys. Rev.}
\newcommand{\CQG}{Class. Quantum Grav.}

\title{Holographic phase transition and quasinormal modes in Lovelock gravity}
\author{Kai Lin$^a$}
\email{lk314159@hotmail.com}
\author{Jeferson de Oliveira$^b$}
\email{jeferson@fisica.ufmt.br}
\author{Elcio Abdalla$^a$}
\email{eabdalla@usp.br}
\affiliation{$^a$Instituto de F\'isica, Universidade de S\~ao Paulo, C.P. 66318, cep 05315-970, S\~ao Paulo, SP, Brazil}
\affiliation{$^b$ Instituto de F\'isica, Universidade Federal de Mato Grosso,\\Cep 78060-900, Cuiab\'a, MT, Brazil }


\date{\today}

\begin{abstract}

In this work we aim at discussing the effects of the higher order curvature terms
on the Lovelock AdS black holes quasinormal spectrum and, in the context of gauge/gravity correspondence, their consequences for the formation of holographic superconductors. We also explore the hydrodynamic limit of the $U(1)$ gauge field perturbations in $d$ dimensions.
\end{abstract}

\maketitle

\section{Introduction}
The AdS/CFT relation, discovered in the framework of string theory \cite{malda,witten}, has surpassed its natural cradle
to spread into the realm of condensed matter by means of the holographic construction \cite{3hs,3hsprl,gubser}. In such a case,
it does not matter much what is the type of the black hole in the AdS space as a physical object, but rather what is
the CFT theory described in the process. Indeed, the CFT theory is the backbone of the construction and classical
perturbations of the gravity set up may lead, changing the black hole, or generally speaking changing the AdS set up,
to valuable information about the CFT counterpart. We
move to new condensed matter systems, thus to new physics.

Recently, a series of models have been considered, with various degrees of success to obtain models concerning condensed
matter systems, see \cite{a,b} for a partial and incomplete list. Several different physical situations have been touched,
such as superconductivity, for perturbations of Reissner-Nordstrom solutions \cite{3hs} and superfluidity
\cite{b}, as well as when dealing with time dependent solutions \cite{bhaseen}, density waves \cite{papa}.
Applications in high energy physics have been particularly important \cite{sonstarinets1},\cite{policastro}. Higher order corrections
to the gravity counterpart have been often used, but a general discussion is still missing \cite{gregkanno,papaetal,bin1,bin2}.

Here, we are going to discuss details of the higher order corrections to gravity and their consequences for the holographic field theory.  
In particular, we consider Lagrangians whose field equations are at most of second order\cite{horndeski} which, in the case of
generalizations of gravity lead us to the Lovelock Lagrangians\cite{lovelock}. The paper is organized as follows. Section \ref{sec:lovelock} provides a review of the $d$-dimensional Lovelock gravity
and the black hole solutions considered in this work.  In Sec. \ref{sec:qnm} we obtain the quasinormal spectrum of $d$-dimensional charged Lovelock black holes due to a scalar probe field. In Secs. \ref{sec:phase_trans} and \ref{sec:r_current} we explore in the probe limit the formation of holographic superconductors in the presence of higher order corrections to the curvature. Also we compute the real time $R$-current correlators due to electromagnetic perturbations due to electromagnetic field. In Sec. \ref{sec:concluding} we discuss the results and some final comments are given.

\section{The Lovelock Gravity}
\label{sec:lovelock}
String theory brought the idea that higher dimensional curvature terms in the gravity action are basically mandatory to
cope with quantum corrections at the Planck scale \cite{stringcorrections}. On the other hand, field equations with higher
time derivatives are unstable. Such a result, originally relying upon Ostrogradsky \cite{ostro} long ago on very general
grounds has been rederived in simple terms \cite{woodard}. Nevertheless, there are theories with complex dynamics involving higher
order terms in the Lagrangian but whose equations of motion are at most second order in time \cite{lovelock,horndeski,mota2012}.
We discuss here the cases of Lovelock gravity as discussed in \cite{cristroncosozanelli}, where, in several space-time dimensions
we have Einstein gravity corrected by higher order terms but with second order differential equations for the fundamental metric
fields. The solutions of the field equations we are considering are those of Refs. \cite{cristroncosozanelli} with
nonvanishing charge, that is, the gravity sector in $d$ dimensions is described by the action
\begin{equation}\label{gravity_sector}
S=-\frac 1{16\pi G}\int d^d x \sum_{p=0}^{k}{\cal L}^{(p)}\quad ,
\end{equation}
where $k$ is an integer strictly smaller than $\frac{d+1}2$ labeling inequivalent theories, the individual Lagrange densities are
\[
{\cal L}^{(p)}=\frac{l^{p-2k}}{d-2k}\left({k \atop p}\right) \epsilon_{\mu_1 \cdot\mu_d}\epsilon^{a_1\cdots a_d}
{\cal R}^{\mu_1\mu_2}_{a_1a_2}\cdots {\cal R}^{\mu_{2p-1}\mu_{2p}}_{a_{2p-1}a_{2p}}e^{\mu_{2p+1}}_{a_{2p+1}}\cdots e^{\mu_d}_{a_d}\quad ,
\]
with $l$ denoting the $d$-dimensional AdS radius related to the cosmological constant $\Lambda$ by
\[
\Lambda=-\frac{(d-1)(d-2)}{2l^2}
\]
and the curvature is
\[
{\cal R}^{\mu\nu}_{ab} =  R^{\mu\nu}_{ab} +\frac 1{l^2}e^{\mu}_ae^{\nu}_b\quad,
\]
where $e^{\mu}_a$ is the vielbein.

It is an established result that the equations of motion are second order in the time derivatives.  The Einstein equations have been
solved \cite{cristroncosozanelli} and a series of AdS black hole solutions emerge from these actions, the most important result used in the present work. Solutions are labeled by the space-time dimension $d$, the integer $k$ defined above.

In order to consider charged solutions, the gravity action (\ref{gravity_sector}) has to be supplemented by the Maxwell action
\begin{equation}\label{gauge_sector}
S_M=-\frac 1{4\epsilon}\int d^dx \sqrt{-g}g^{\mu\rho}g^{\nu\sigma} F_{\mu\nu}F_{\rho\sigma}\quad,
\end{equation}
where $\epsilon$ is related to the vacuum permeability $\epsilon_0$.

Solutions to the Einstein-Maxwell system are labeled by the space-time dimension $d$, the integer $k$ defined above and the charge $Q$. From \cite{cristroncosozanelli,arostroncosozanelli}, those solutions read
\begin{eqnarray}\label{metric1}
ds^2=-\left( \eta +\frac{r^2}{l^2} -g_k (r)\right) dt^2 + \frac{dr^2}{\eta+\frac{r^2}{l^2} -g_k (r) }
+ r^2 d\Sigma_{d-2}^2\quad ,
\end{eqnarray}
where $\eta=-1,0,1$ defines the topology. For $\eta=1$, $d\Sigma^2_{d-2}$ is the angular measure on the sphere, $\eta=0$ implies a flat black hole where
$d\Sigma^2_{d-2}$ is the flat metric and $\eta=-1$ corresponds to the hyperbolic metric. The metric depends on the charge $Q$ of the black hole by means of the expression
\begin{equation}
g_k = \left( \frac{2\hat G M + \delta_{d-2k,1}}{r^{d-2k-1}}-\frac{\epsilon\hat G}{d-3}\frac {Q^2}{r^{2(d-k-2)}} \right)^{1/k}\quad ,
\end{equation}
where $\hat{G}$ is the gravitational constant, $\delta_{d-2k,1}$ is
the standard Kronecker delta and $M$ is the black hole mass. For a
generic value of $k$, the black holes described by the line element
(\ref{metric1}) have two horizons ($r_{-}$,$r_{+}$) located at the
zeros of $g_{tt}$, satisfying $r_{-}<r_{+}$. The black hole family
describe by (\ref{metric1}) include the $d$-dimensional
Reissner-Nordstr\"om AdS-black holes for $k=1$ and the charged
AdS-Gauss-Bonnet black holes for $k=2$ and $d>5$.  For a discussion on
the causal structure of Gauss-Bonnet  gravity, see \cite{izumi}. Also, the line element (\ref{metric1})
is asymptotically AdS for all values of $k$ and $d$.




\section{Lovelock Scalar quasinormal modes}\label{sec:qnm}
In this section we are going to explore the quasinormal spectrum of charged AdS-Lovelock black holes considering a probe scalar field evolving at such geometry.

The black hole quasinormal modes (QNM) of asymptotically AdS black
holes is obtained by considering probe fields dynamics supplemented by
ingoing boundary conditions at the event horizon and Dirichlet
boundary conditions at spatial infinity \cite{lemos1,lemos2}. In the context of AdS/CFT correspondence the decay of QNM is interpreted as the return to equilibrium of a thermal state in the quantum field theory at finite temperature living at the AdS boundary \cite{HH}. For a recent review on the subject see \cite{berti}, \cite{konoplya} and the references therein. In particular, scalar fluctuations on the bulk geometry are related to the poles of thermal retarded Green function \cite{sonstarinets} and the electromagnetic perturbations are associated to the poles of retarded Green functions of $R-$symmetry currents at the boundary.

The next procedure is standard but new. We consider the scalar perturbations of the above system. Scalar perturbations are easily obtained. We rewrite the metric in a form that we use in the numerical analysis below, that is,
\begin{eqnarray}
ds^2=-f(r) dt^2 + \frac{dr^2}{f(r) } + r^2 d\Sigma_{d-2}^2\quad ,\label{metric}
\end{eqnarray}
with
\begin{eqnarray}
f(r)=\eta
+\frac{r^2}{l^2}-\left[\frac{M}{r^{d-2k-1}}-\frac{Q^2}{r^{2(d-k-2)}}
\right]^{\frac 1k} \quad .
\end{eqnarray}
Depending on the curvature (thus on $\eta$), the angular part of
(\ref{metric}) changes accordingly. We can rewrite the parameters in
terms of the inner horizon $r_{-}$ and the event horizon $r_{+}$ as
\begin{eqnarray}
M&=&\frac{1}{r_{+}^{d-3}-r_{-}^{d-3}}\left[r_{+}^{2d-4-2k}\left(\eta+\frac{r_{+}^2}{l^2}\right)^k-r_{-}^{2d-4-2k}
\left(\eta+\frac{r_{-}^2}{l^2}\right)^k\right],\nb\\
Q^2&=&\frac{1}{r_{-}^{3-d}-r_{+}^{3-d}}\left[r_{+}^{d-1-2k}\left(\eta+\frac{r_{+}^2}{l^2}\right)^k-r_{-}^{d-1-2k}
\left(\eta+\frac{r_{-}^2}{l^2}\right)^k\right].
\end{eqnarray}

In this paper, we study the planar black hole case, namely $\eta=0$, and
without loss the generality, we set $l=1$.
Now, it is standard to compute the scalar modes. Because it is an
anti-de Sitter spacetime, we should use Horowitz-Hubeny method
\cite{HH} to calculate the scalar quasinormal modes of this 
black hole.
According to this method, we set $v=t+\int\frac{dr}{f(r)}$, so the
metric is rewritten as
\begin{eqnarray}
ds^2=-f(r) dv^2 + 2dv dr + r^2 d\Sigma_{d-2}^2\quad ,
\label{metric2}
\end{eqnarray}
and then the scalar equation is given by
\begin{eqnarray}
f(r)\phi''+(f'-2i\omega)\phi'-V(r)\phi=0,
\label{quasinormalequation}
\end{eqnarray}
where
$V=\frac{(d-2)f'}{2r}+(d-4)(d-2)\frac{f}{4r^2}+(d-2)\frac{L^2}{r^2}$.
The transformation $z=\frac{1}{r}$ is introduced, so that the region
of variable becomes $0\leq z\leq h$ ($h=\frac{1}{r_+}$). The
boundary condition at event horizon require $\phi(r_+)=1$ but $\phi$
should vanish at infinity. Therefore, the scalar field equation is
given by
\begin{eqnarray}
s(z)\frac{d^2\phi}{dz^2}+\frac{t(z)}{z-h}\frac{d\phi}{dz}+\frac{u(z)\phi}{(z-h)^2}=0,
\label{quasinormalequation1}
\end{eqnarray}
where
\begin{eqnarray}
s(z)&=&-\frac{z^4f}{z-h},\nb\\
t(z)&=&-z^{2}\left(z^{-2}\frac{df}{dz}+2zf+2i\omega\right),\nb\\
u(z)&=&(z-h)V. \label{quasinormalequation2}
\end{eqnarray}
we can expand $s(z)=\sum s_i(x-h)^i$, $t(z)=\sum t_i(x-h)^i$
and $u(z)=\sum u_i(x-h)^i$ and $\phi(z)=\sum a_i(x-h)^i$ at event
horizon $z=h$. Considering the boundary condition at horizon, we
have $a_0=1$, and substitute $s$, $u$, $t$ and $\phi$ into
Eq.(\ref{quasinormalequation2}), we obtain the recursion relation
\begin{eqnarray}
a_n=-\frac{1}{P_n}\sum\limits_{i=0}^{n-1}\left[i(i-1)s_{n-i}+it_{n-i}+u_{n-i}\right]a_i.
\label{quasinormalequation3}
\end{eqnarray}
where $P_n=n(n-1)s_0+nt_0$, so all the $a_i$ can be obtained.
Finally, according to another boundary condition $\phi(0)=0$, we
always can get the value of $\omega$ from equation $\sum\limits_{i}
a_i=0$. It is very convenient to use {\it{Wolfram Mathematica}} to realize the above
process, so we use this software to calculate the quasinormal
modes of this black hole.  We find a sequence of quasinormal
frequencies as function of the temperature of the black hole. It is
a tedious but straightforward procedure. Nevertheless, we find some
interesting results.

\begin{figure}[htb!]
\centering
\includegraphics[scale=0.7]{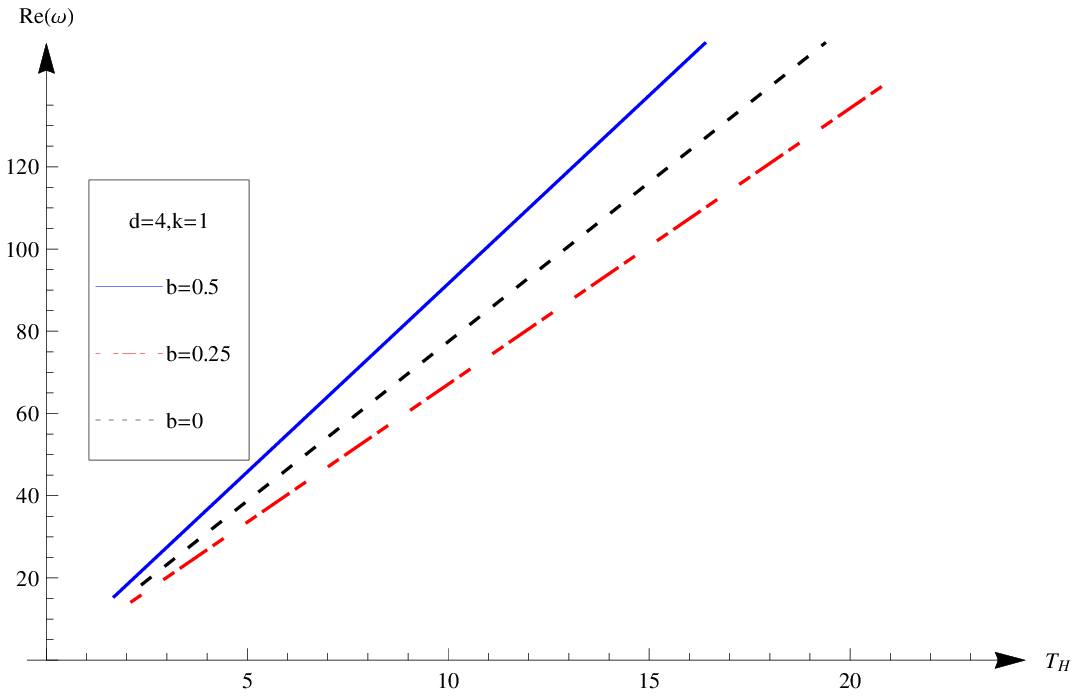}
\includegraphics[scale=0.6]{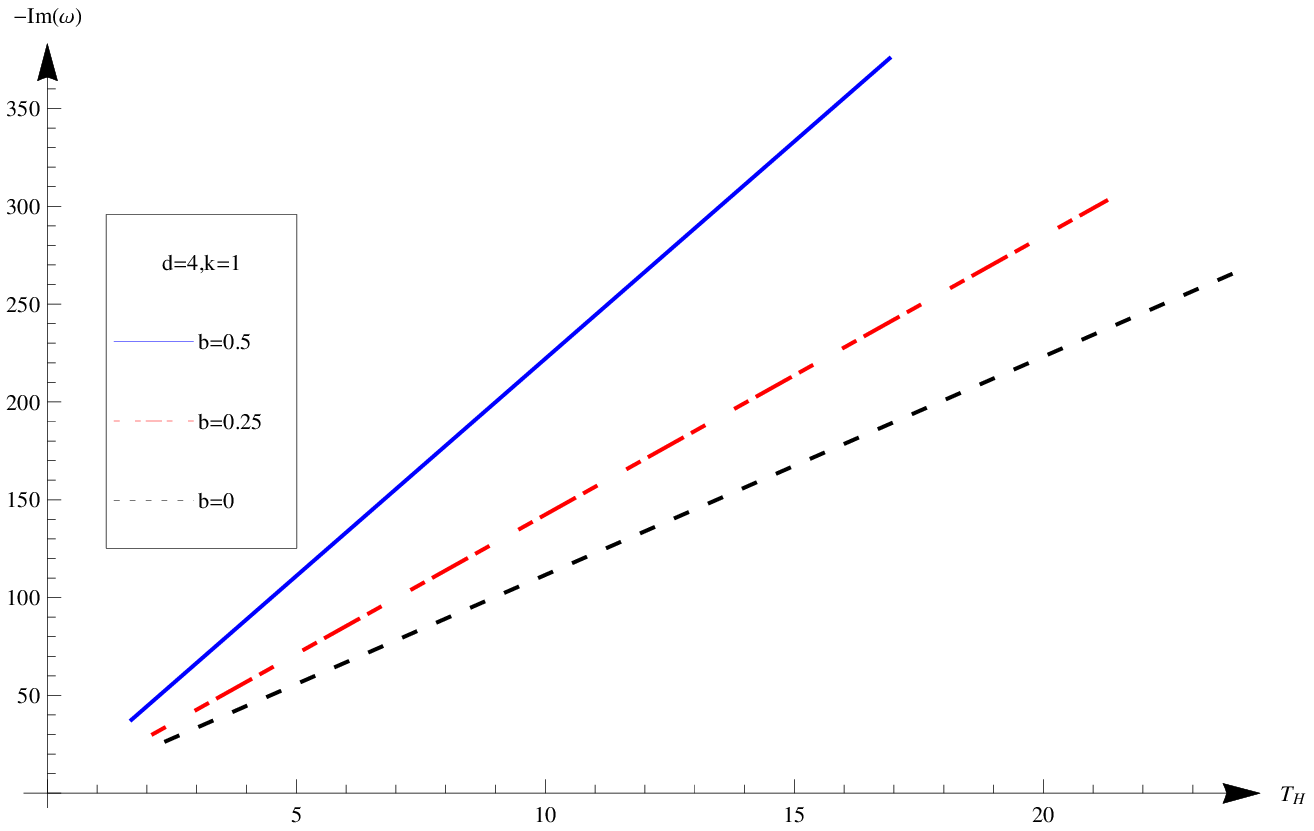}
\caption{Real(left) and imaginary(right) quasinormal modes behavior in terms of the Hawking temperature for $d=4, k=1$. }
\label{variousdim4}
\end{figure}

\begin{figure}[htb!]
\centering
\includegraphics[scale=0.85]{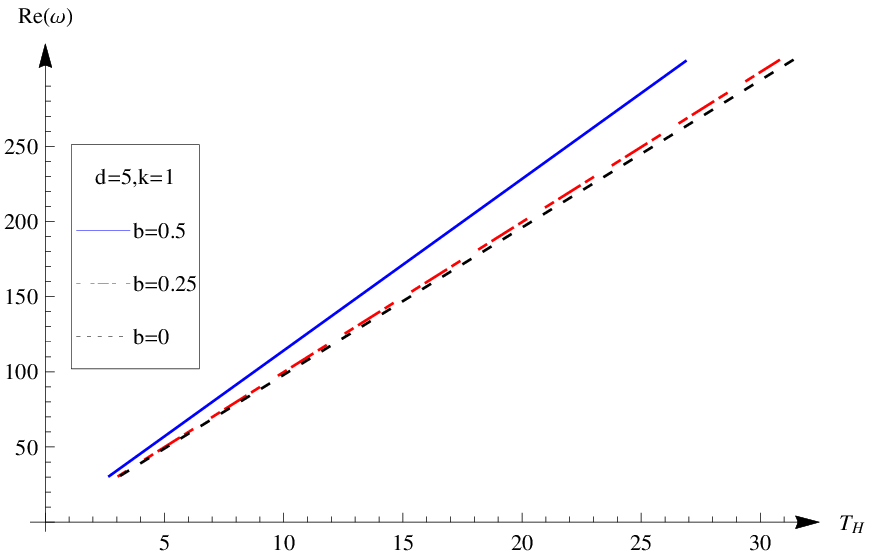}
\includegraphics[scale=0.85]{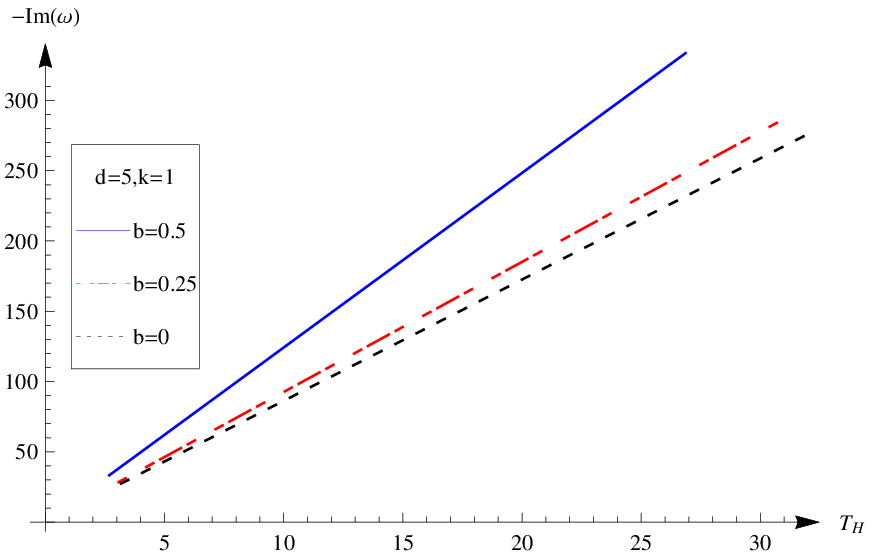}\\
\caption{Real(left) and imaginary(right) quasinormal modes behavior in terms of the Hawking temperature for $d=5, k=1$. }
\label{variousdim5}
\end{figure}

\begin{figure}[htb!]
\centering
\includegraphics[scale=0.5]{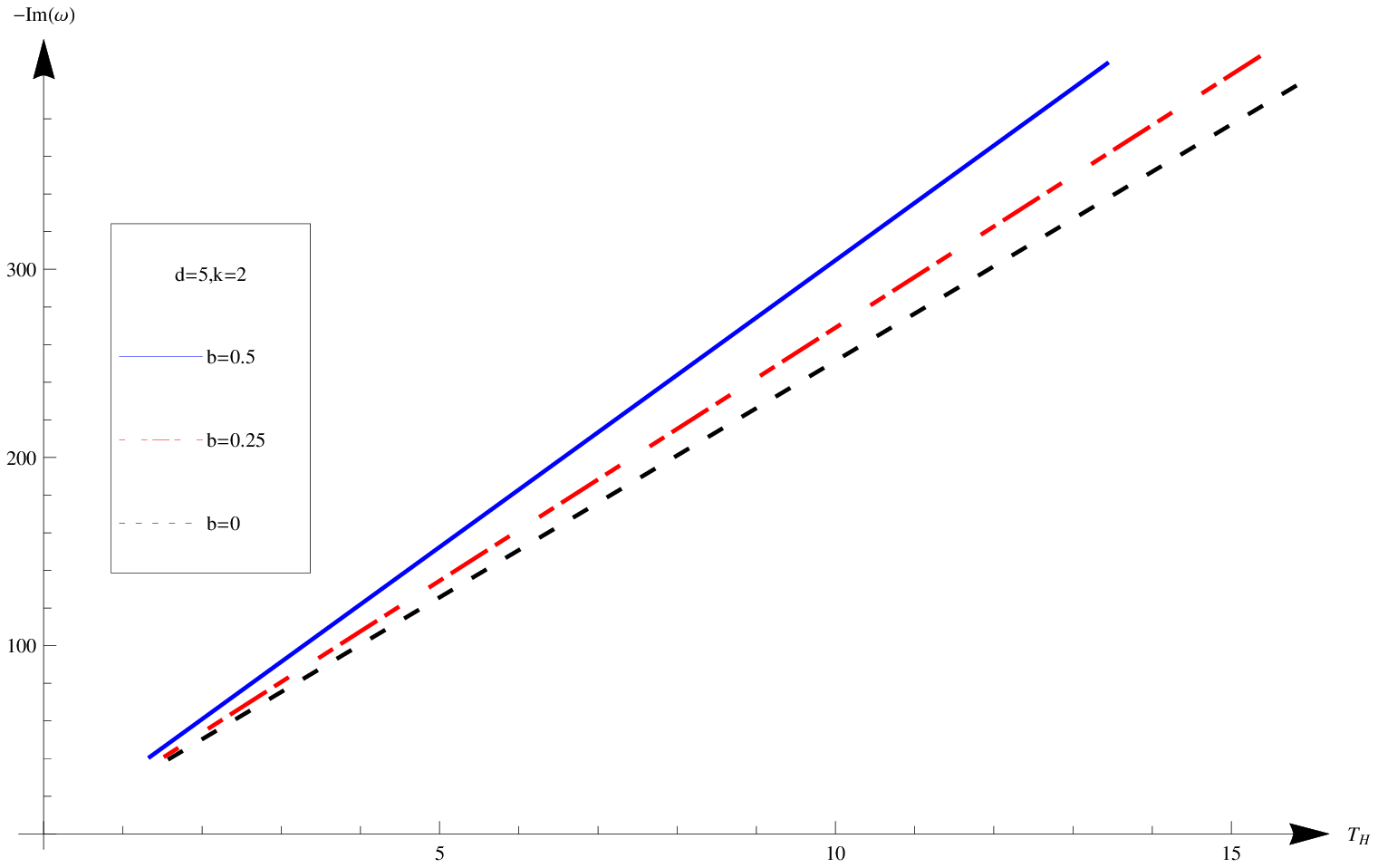}
\includegraphics[scale=0.6]{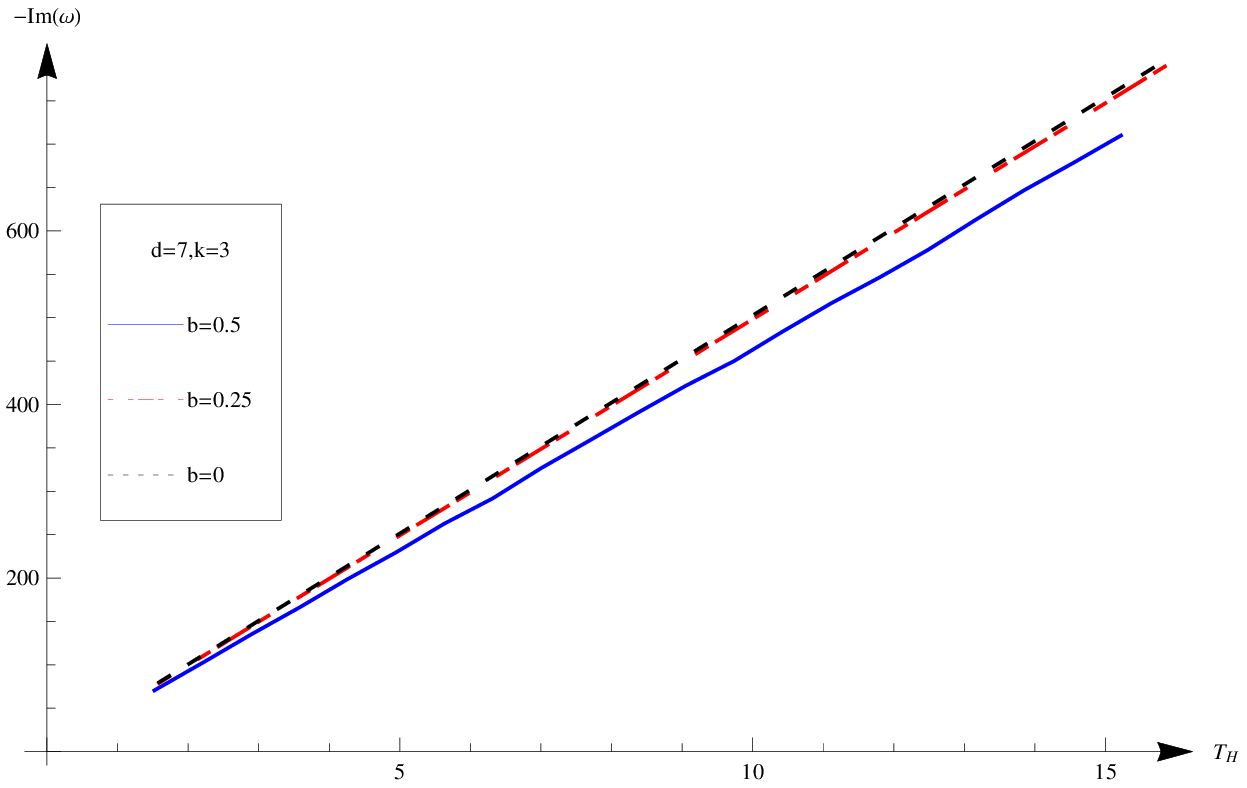}\\
\caption{Purely damped quasinormal modes behavior in terms of the
  Hawking temperature for $d=5, k=2$ and $d=7, k=3$ }
\label{variousdim52}
\end{figure}

\begin{figure}[htb!]
\centering
\includegraphics[scale=0.8]{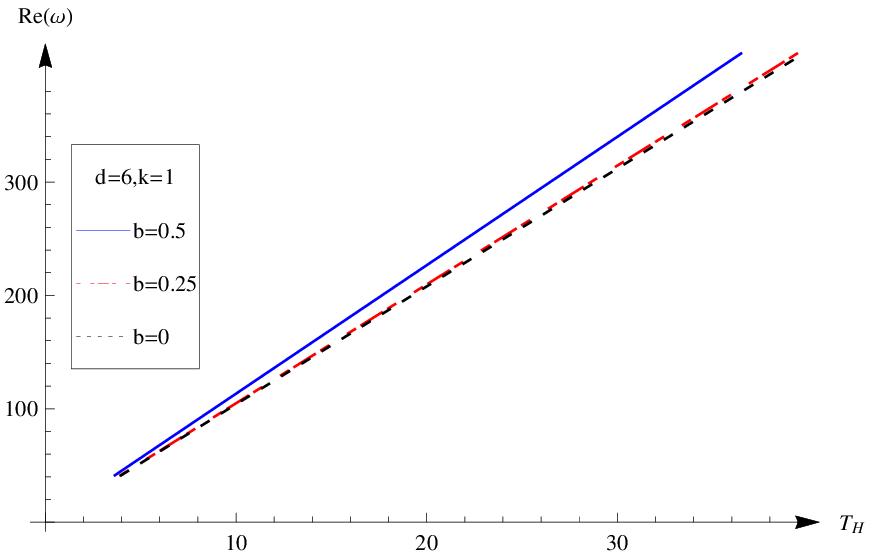}
\includegraphics[scale=0.8]{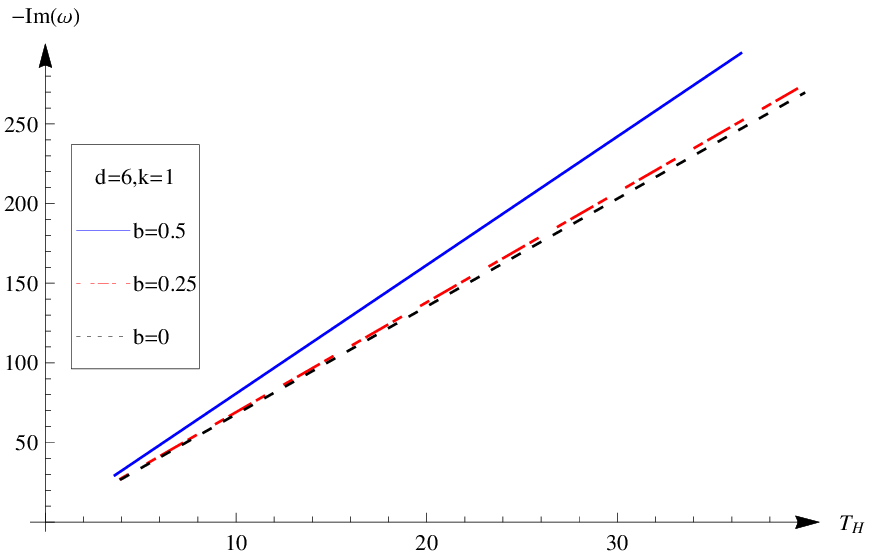}\\
\includegraphics[scale=0.8]{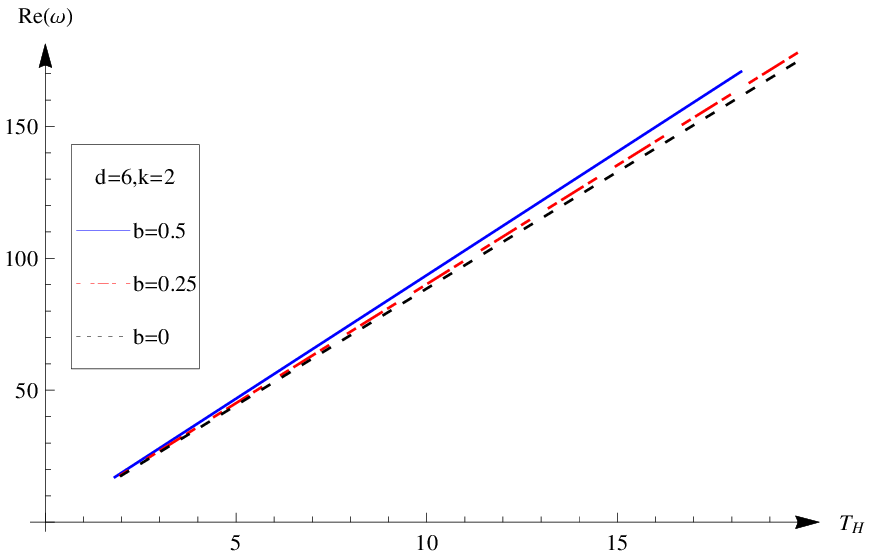}
\includegraphics[scale=0.8]{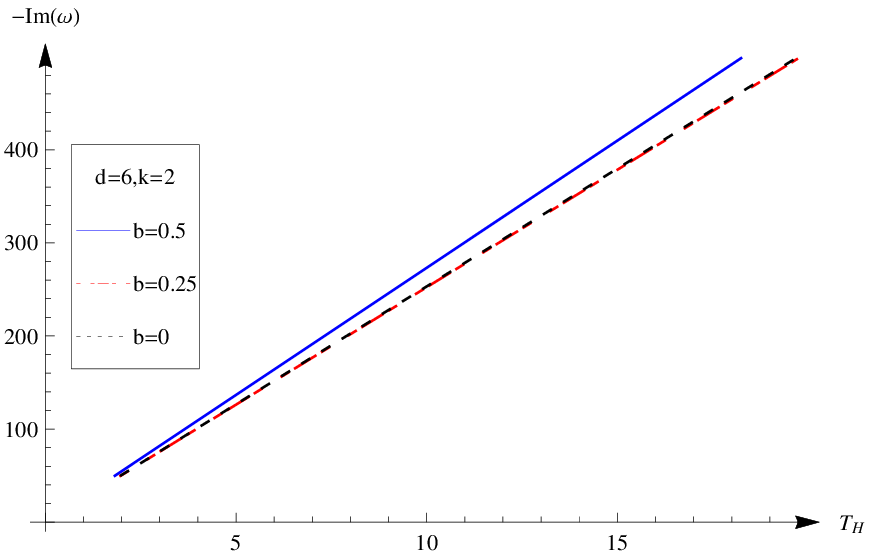}\\
\caption{Effect on the real part(left) and imaginary part(right)of quasinormal frequencies by adding charge to the six-dimensional case for $k=1$, $k=2$. }
\label{variousdim62}
\end{figure}

The first noteworthy result with angular quantum number $l=0$  is
the fact that spaces with higher values of $k$ are stiffer, namely
have larger values for the imaginary part of the frequency, as shown
in Figs. \ref{variousdim4} through \ref{variousdim62}. Also, we
clearly see that the real and imaginary part of frequencies scales
linearly with the Hawking temperature, which is expected of AdS black
holes \cite{HH}. We have the results for various values of the
temperature as given in Table I, where $b$ is the value of black
hole charge $Q$.

\begin{table}[ht]
\caption{\label{Table} Various values of quasinormal frequencies, where $T$ is the Hawking temperature}
\centering
\begin{tabular}{c c c c c c c}
         \hline\hline
$b$&    $d=4$&  $d=5$, $k=1$&   $d=5$, $k=2$&   $d=6$, $k=1$    & $d=6$, $k=2$
        \\
        \hline
0&$ (7.747 - 11.158 i) T $&  $(9.800 -8.620i)T$  & $(-25.133 i)T$ &$ (10.394 - 6.769 i) T  $ &$  (8.854 - 25.320 i) T $
          \\
0.25& $ (6.712 - 14.240 i) T  $ &    $ (9.981- 9.257i)T$ & $(-26.899 i)T$&$ (10.483 - 6.902 i) T  $ &  $ (9.021 - 25.235 i) T  $
          \\
0.5 & $ (9.156 - 22.211 i) T  $ &  $(11.416-12.4206i)T $ &    $(-30.480i)T$ & $ (11.33 - 8.069 i) T  $ &  $(9.361 - 27.322 i) T $
          \\
        \hline
\end{tabular}
\end{table}

As we see in Fig. \ref{variousdim52}, the scalar quasinormal
frequencies for a five and seven-dimensional Lovelock black hole, with
$k=2$ and $k=3$, respectively, are purely damped, namely there is no
oscillatory phase for the perturbation. Such a result seems to be a
general feature of Lovelock theories with $d=2k+1$ since the gravity
theory reduces to Chern-Simons gravity in these cases. The purely
damped frequencies are  not new in literature.  The same result has
been found in the behavior of scalar quasinormal modes of the three
dimensional Lifshitz black hole \cite{berthajefcep}, whose gravity
theory is the new massive gravity (NMG). The corresponding action, as in the
Lovelock case, contains higher order corrections in the curvature. The
dynamics of probe scalar fields in higher dimensional Lifshitz black
holes (d=5,...,10) do not show an oscillating phase either \cite{owenelciofideljef}. Thus, at least in
case of Lovelock gravity with $d=2k+1$ and Lifshitz black holes, the
 purely damped modes are related to the higher curvature terms.

Also, we observe that the effect of adding charge to the  black hole is to increase the quasinormal frequency value. There is also a quite important increase in the imaginary part of the frequencies for $k=2$, when the models seems to get stiffer. This effect is less pronounced for the six-dimensional case, see Fig. \ref{variousdim62}.

\section{The Phase Transition and Conductivity}\label{sec:phase_trans}

According to the AdS/CFT dictionary, the scalar perturbation $\psi$
corresponds, at the AdS border, to the order parameter of the
conformal field theory. The gauge field perturbation gives rise to
the border source and to the current. We can thus analyze
whether we can have a superconducting phase and compute the
conductivity. As it turns out, we have the conductivity as a
function of the frequency, what is a physically relevant object to
study the properties of the conformal field theory at the border (or
else, of the condensed matter system at the border). For later
purposes it is going to be useful to write the above function in
terms of the event horizon radius $r_+$ and the Cauchy horizon
$r_c\equiv \gamma r_+$. From this point on we shall work on flat topology,
$\eta=0$ and $l=1$. We have 
\begin{eqnarray}
M= \frac{\gamma^{2d-4}-1}{b^{d-3}-1}r_+^{d-1} ,\hspace{0.3cm}
Q^2 = \frac{\gamma^{2d-4}-\gamma^{d-3}}{\gamma^{d-3}-1}r_+^{2d-4} \quad ,
\end{eqnarray}
where $\gamma=\frac{r_-}{r_+}$. In terms of the new parameters we have
\begin{equation}
f(r)=r^2 -
r_+^{\frac{d-1}{k}}r^{2(1+\frac{2-d}{k})}
\left[\frac{1-\gamma^{2d-4}}{1-\gamma^{d-3}}r^{d-3}-\gamma^{d-3}\frac{1-\gamma^{d-1}}{1-\gamma^{d-3}}\right]^{\frac{1}{k}}\quad ,
\end{equation}
and the Hawking temperature is given in terms of the local gravity at
the black hole event horizon,
\begin{equation}
T_c= \frac{d-1-\gamma^{d-3}[2d-4-\gamma^{d-1}(d-3)]}{4k(1-\gamma^{d-3})\pi}r_+.
\end{equation}

In order to obtain the phase transition, we consider the Lovelock gravity action (\ref{gravity_sector}) coupled to a classical charged scalar field $\psi$ and the electromagnetic gauge field $A_{\mu}$, whose action is

\begin{equation}
\label{higgs-type}
S_{c}=\int d^{d}x\sqrt{-g}\left[-\frac{1}{4\epsilon}F_{\mu\nu}F^{\mu\nu}-|\nabla\psi - i q A\psi |^{2}-m^{2}|\psi |^{2}\right],
\end{equation}
where $\nabla$ is the covariant derivative, $q$ and $m$ are the scalar field charge and mass respectively.

We consider the equation of motion of the matter and gauge fields,
in such a way that scalars are functions of the radial variable in
order to define an order parameter at the border. The scalar
potential corresponding to the gauge field ($\phi\equiv {\cal A}_0$)
is, consequently, a function of the radial variable. The vectorial
components of the gauge field are functions of $t$ and $r$. Without
loss of generality we consider only ${\cal A}_x$, whose time
dependence is harmonic, that is, ${\cal A}_x(\vec x,t)={\cal
A}_x(\vec x)e^{-i\omega t}$.

We are going to consider the equations of motion of the electric
potential $A_0=\Phi(r)$, of the scalar field $\Psi(r)$ and the
$x$-component of the vector potential, ${\cal A}_x(r)$. Moreover, it
is useful to change variables from $r$ to $z=\frac 1r$. Also,
foreseeing the asymptotic behavior of the fields, we redefine them
as $\phi (z) = \Phi (1/z)$, $\psi(z) = \Psi (1/z) z^{-\lambda_f}$,
$A_x(z)={\cal A}_x(1/z)$. We in this paper use the shooting method to
calculate numerically the holographic superconductor and the
conductivity. The Maxwell-Klein Gordon equations in the probe limit
for $k=1$ and $d=5$ read 
\begin{eqnarray}
&&\psi_{5,1}''+\frac{3(\gamma^4+b^2)z^6-(\gamma^4+b^2+1)z^4-3}{(z^3-z)[(\gamma^4+\gamma^2)z^4-z^2-1]}\psi_{5,1}'+\frac{m^2(1-z^2)[(\gamma^4+\gamma^2)z^4-z^2-1]+z^2\phi_{5,1}}{(z^3-z)^2[(\gamma^4+\gamma^2)z^4-z^2-1]^2}\psi_{5,1}=0,\nonumber\\
&&\phi_{5,1}''-\frac{\psi_{5,1}'}{z}-\frac{2\psi_{5,1}^2\phi_{5,1}}{z^2(z^2-1)[(\gamma^4+\gamma^2)z^4-z^2-1]}=0,\nonumber\\
&&A_{x,5,1}''+\frac{5(\gamma^4+\gamma^2)z^6-3(\gamma^4+\gamma^2+1)z^4-1}{(z^3-z)[(\gamma^4+\gamma^2)z^4-z^2-1]}A_{x,5,1}'+\frac{z^2\omega^2-2(z^2-1)[(\gamma^4+\gamma^2)z^4-z^2-1]\psi_{5,1}^2}{(z^3-z)^2[(\gamma^4+\gamma^2)z^4-z^2-1]^2}A_{x,5,1}\nonumber\\
&&=0.\nonumber
\end{eqnarray}

By the shooting method,  we choose the
value of the fields near the horizon, solve the differential
equations to the spatial infinity and compare with the boundary
condition. For solving the differential equation we choose the
functions as power series of $z-1$. We thus obtain the value of the
function at the boundary and compare with the boundary condition. We
subsequently consider the cases $d=5,k=2$, $d=6,k=1$ and $d=6,k=2$,
whose equations of motion are given in the appendix.

\subsection{Numerical analysis }
\label{analysis}

Let us first concentrate on the 5-dimensional case, where the function $f(r)$ defining the metric is given by
\begin{eqnarray}
f(r) &=& \eta +\frac {r^2}{R^2} -\frac{r_e^{\frac{4}{k}-2}
r^{2-\frac{6}{k}}}{(b^2-1)^{\frac{1}{k}}b^2}\left[b^6(r^2-r_e^2)
\left(\frac{b^2 r_e^2}{R^2}+\eta
\right)^k-b^{2k}(r^2-r_e^2b^2)\left(\frac{r_e^2}{R^2}+\eta
\right)^k\right]^{\frac{1}{k}} .
\end{eqnarray}
Here, $\eta =+1,0,-1$ depending on whether the solution has
positive, zero or negative curvature. The parameter $b$ is a measure
of the charge of the black hole, $R$ is the inverse of the
cosmological constant and $r_e$ the event horizon.

We search for static solutions for the electric potential and for
the scalar field seeking at the order parameter at the border.
Moreover, we look for a vector potential at a given frequency (as
above) in order to test the Ohm's law. The fields obey the coupled
differential equations
\begin{eqnarray}
&&\psi^{\prime\prime}(r)+\left(\frac{f^\prime(r)}{f(r)}+\frac{d-2}{r}
\right)\psi^\prime+\left\lbrack\frac{\phi^2(r)}{f^2(r)}
-\frac{m^2}{f(r)}\right\rbrack \psi(r)=0,\\
&&\phi^{\prime\prime}+\frac{d-2}{r}\phi^\prime -\frac{2q^2\psi^2(r)}{f(r)}\phi(r)=0,\\\label{cond}
&&A_x^{\prime\prime}(r)+\left(\frac{f^\prime(r)}{f(r)}+\frac{d-4}{r}
\right)A_x^\prime(r)+\left(\frac{\omega^2}{f^2(r)}-\frac{2\psi^2(r)}{f(r)}\right)A_{x}(r)=0.
\end{eqnarray}

\subsection{Results for phase transition}
\label{sec:result_phase}

According to the usual AdS/CFT dictionary, when we approach the AdS boundary the expansion of the perturbations near the boundary leads to CFT fields with well-defined physical interpretation \cite{witten,3hs}. For the scalar field, in particular, we have the expansion
\begin{equation}
\psi(r)=\psi^{(1)}\frac 1r+\psi^{(2)}\frac 1{r^2} +{\hbox{\it{higher order in }} \left(\frac 1r\right)}\quad .
\end{equation}
The expansion coefficients $\langle{\cal O}_i\rangle=\sqrt 2 \psi^{(i)}$ are, according to the above mentioned dictionary, order parameters of the boundary theory, as long as we choose appropriated boundary conditions, that is, if $\psi^{(1)}=0$ we define $\langle{\cal O}_2\rangle$ and for $\psi^{(2)}=0$ we define $\langle{\cal O}_1\rangle$  . The computaion of either field uses the shooting method found by \cite{3hsprl}.

\begin{figure}[htb!]
\centering
\includegraphics[scale=0.6]{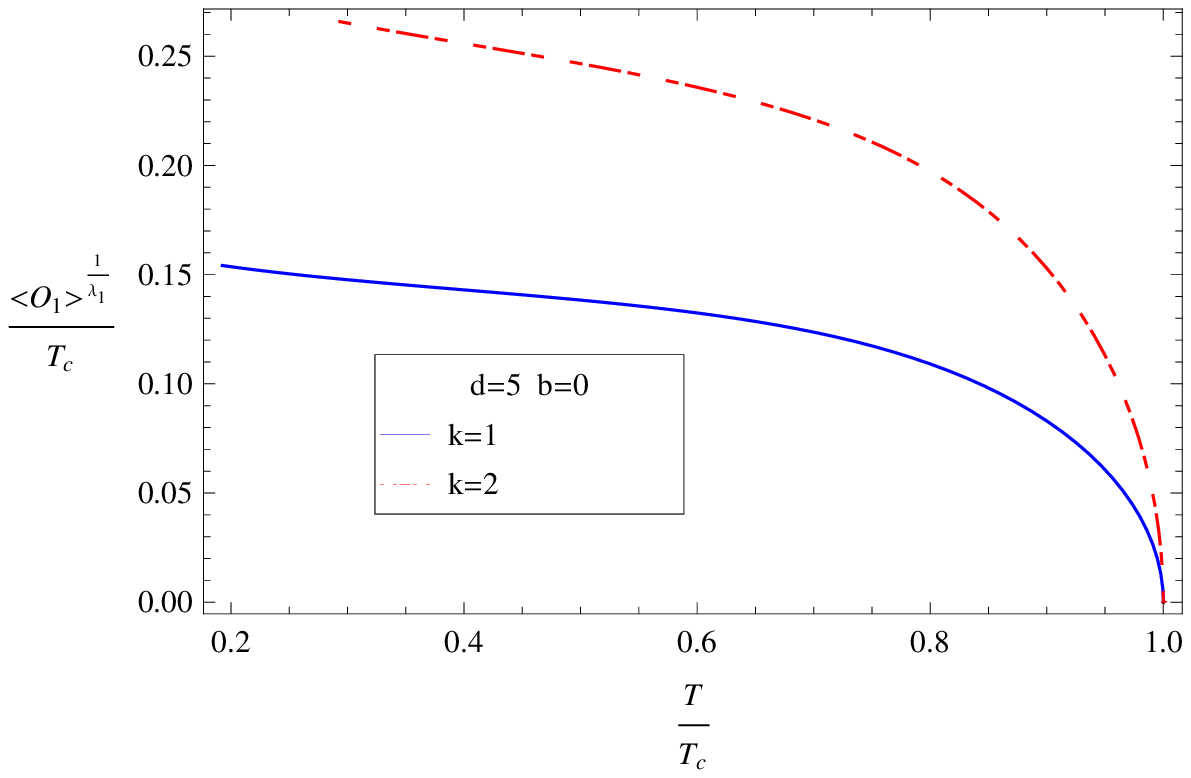}
\includegraphics[scale=0.6]{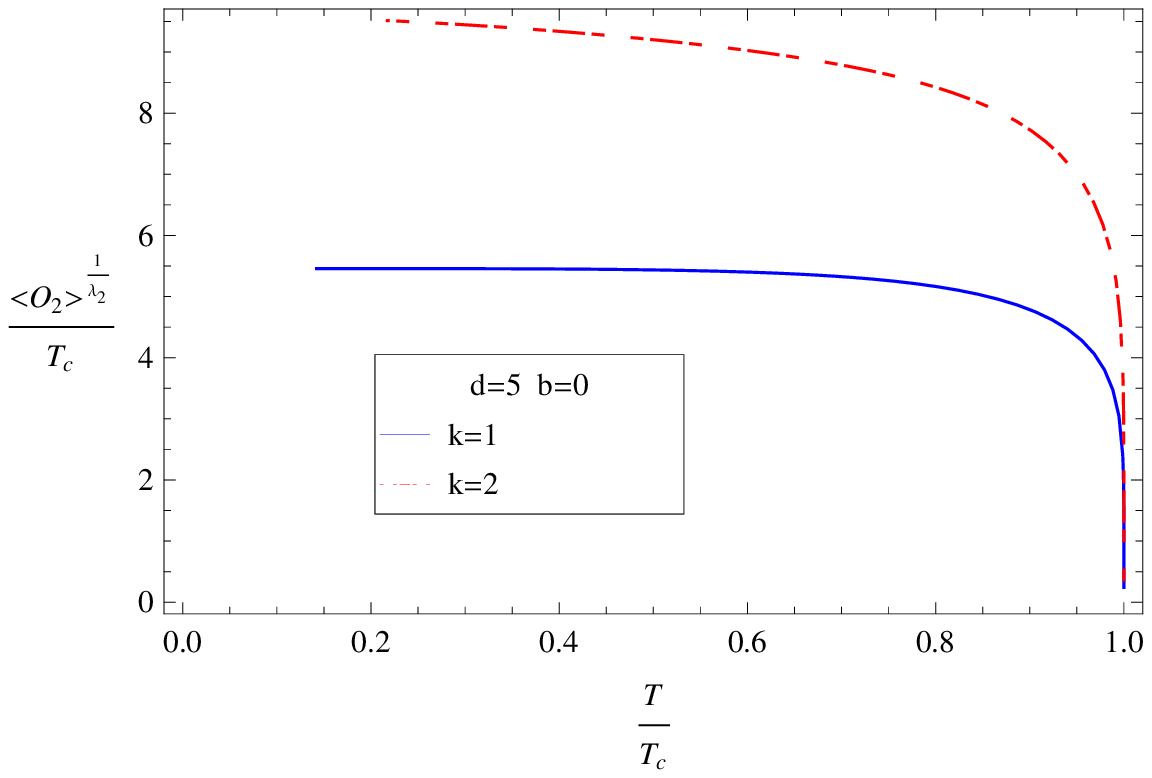}
\caption{Condensation of operators $\langle\mathcal{O}_{1}\rangle$ and $\langle\mathcal{O}_{1}\rangle$ for the five-dimensional uncharged case ($b=0$).}
\label{fixd5_condensation_charge0}
\end{figure}

\begin{figure}[htb!]
\centering
\includegraphics[scale=0.6]{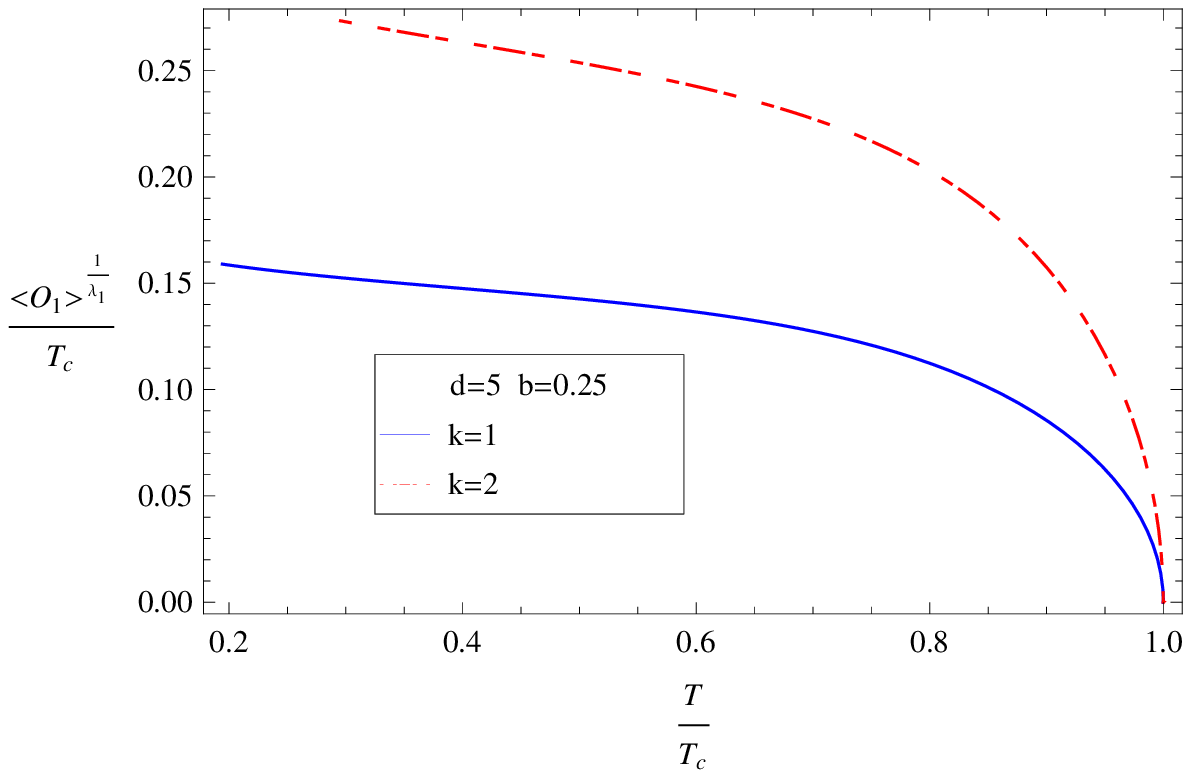}
\includegraphics[scale=0.6]{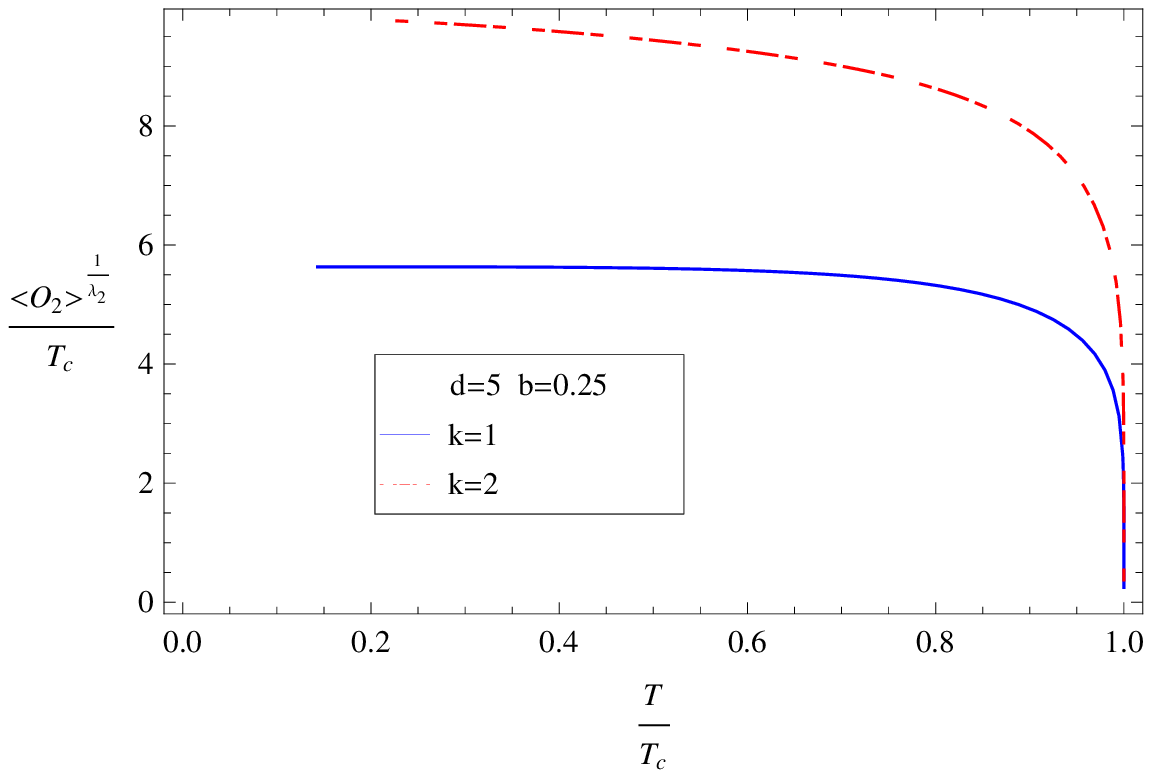}
\caption{Condensation of operators $\langle\mathcal{O}_{1}\rangle$ and $\langle\mathcal{O}_{1}\rangle$ for the five-dimensional charged case ($b=0.25$).}
\label{fixd5_condensation_charge025}
\end{figure}

We considered various choice of parameters. Generally speaking, the order parameter $\langle{\cal O}_2\rangle$ is larger than
$\langle{\cal O}_1\rangle$, and the one corresponding to $k=2$ larger than the one corresponding to $k=1$, see Fig. \ref{fixd5_condensation_charge0} and \ref{fixd5_condensation_charge025}. This result about $k$ means
that the nonlinearity enhances the order parameter, but strangely
enough the conductivity goes the other way (see next subsection),
namely the conductivity (both real and imaginary part) are smaller for
$k=2$. Thus, order does not mean, in this case, better conductivity
properties.

\begin{figure}[htb!]
\centering
\includegraphics[scale=0.6]{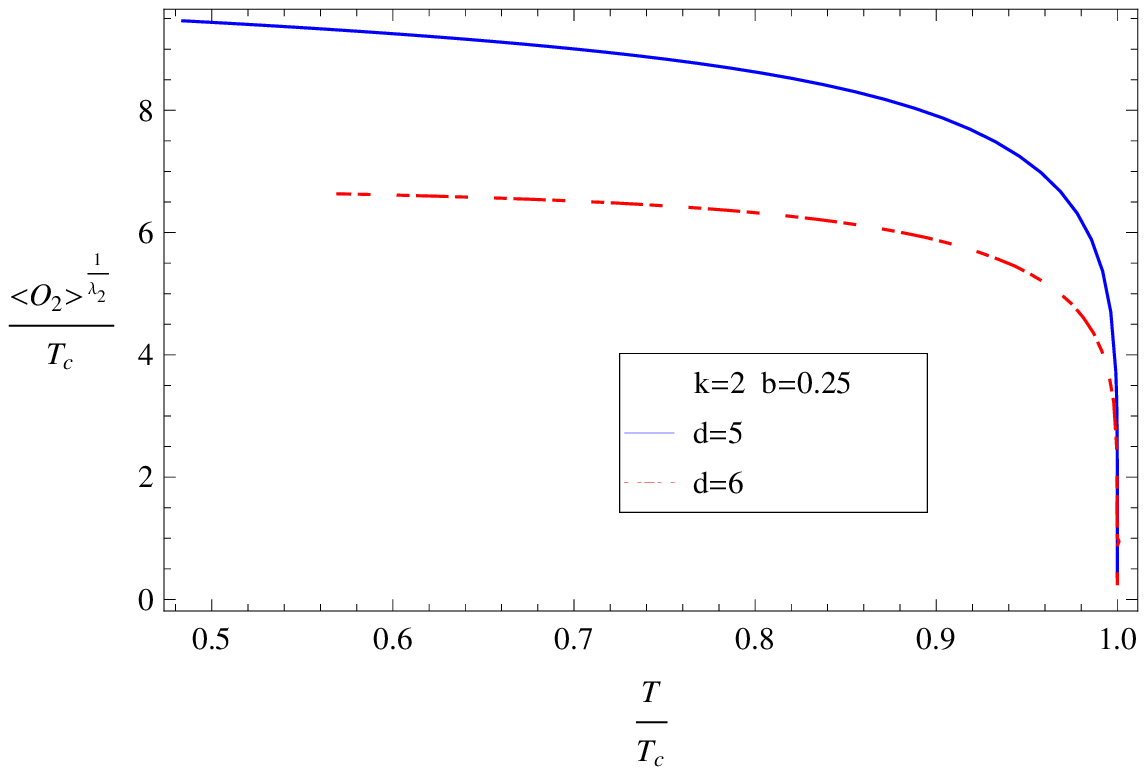}
\includegraphics[scale=0.6]{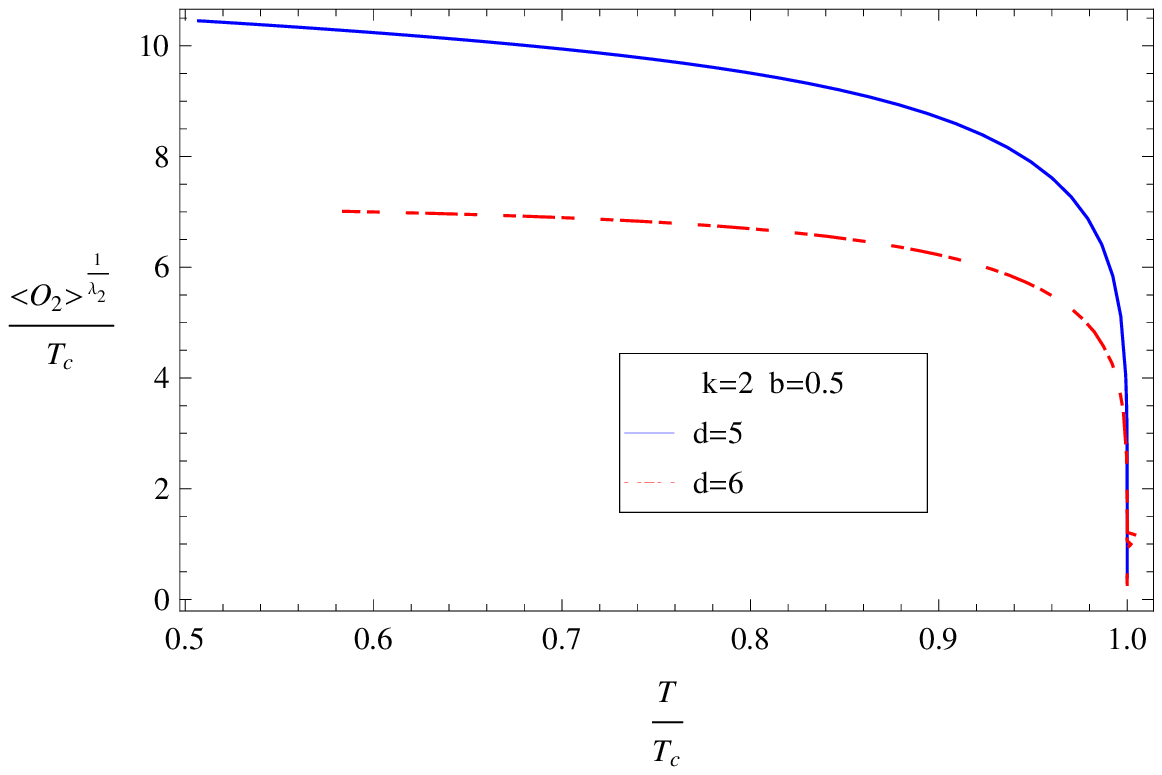}
\caption{Condensation dependence on the dimensionality and the values of $k$ }
\label{varying_d}
\end{figure}

The effect of dimensionality upon the phase transition is to lower the
value of the condensate as the number of spatial dimensions
increase. Such an effect is present in both condensates $\langle{\cal
  O}_1\rangle$ and $\langle{\cal O}_2\rangle$, see Fig. \ref{varying_d} for
an example.

\subsection{Results for conductivity}
\label{sec:result_cond}

Now, we are going to compute the conductivity for each boundary operator $\langle\mathcal{O}_{1}\rangle$ and $\langle\mathcal{O}_{2}\rangle$ following the standar AdS/CFT recipe \cite{3hs}. Solving numerically the equation (\ref{cond}), imposing ingoing wave boundary conditions at the black hole event horizon and considering the asymptotic behavior of $A_{x}$ for large $r$, we have that the leading term is the current $\langle J_{\mu}\rangle$ and the sub leading one the dual source $A_{x}^{(0)}$, both defined at the AdS border.

\begin{figure}[htb!]
\centering
\includegraphics[scale=0.6]{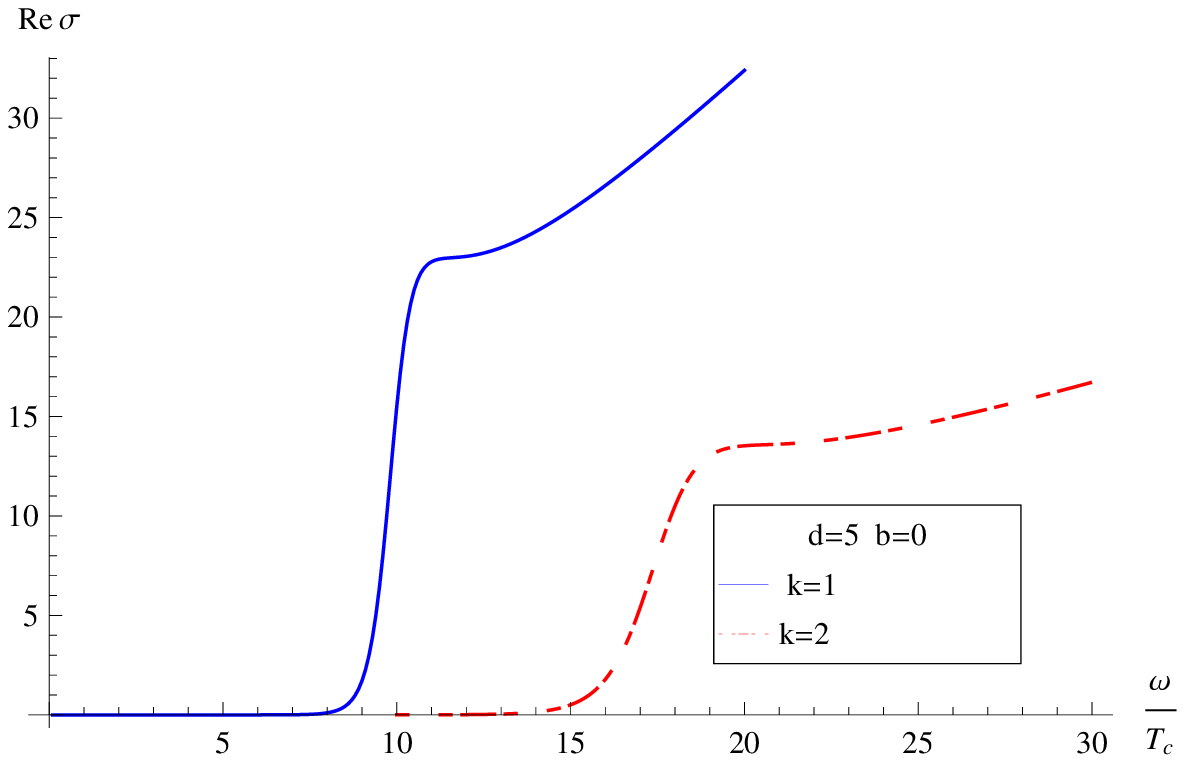}
\includegraphics[scale=0.6]{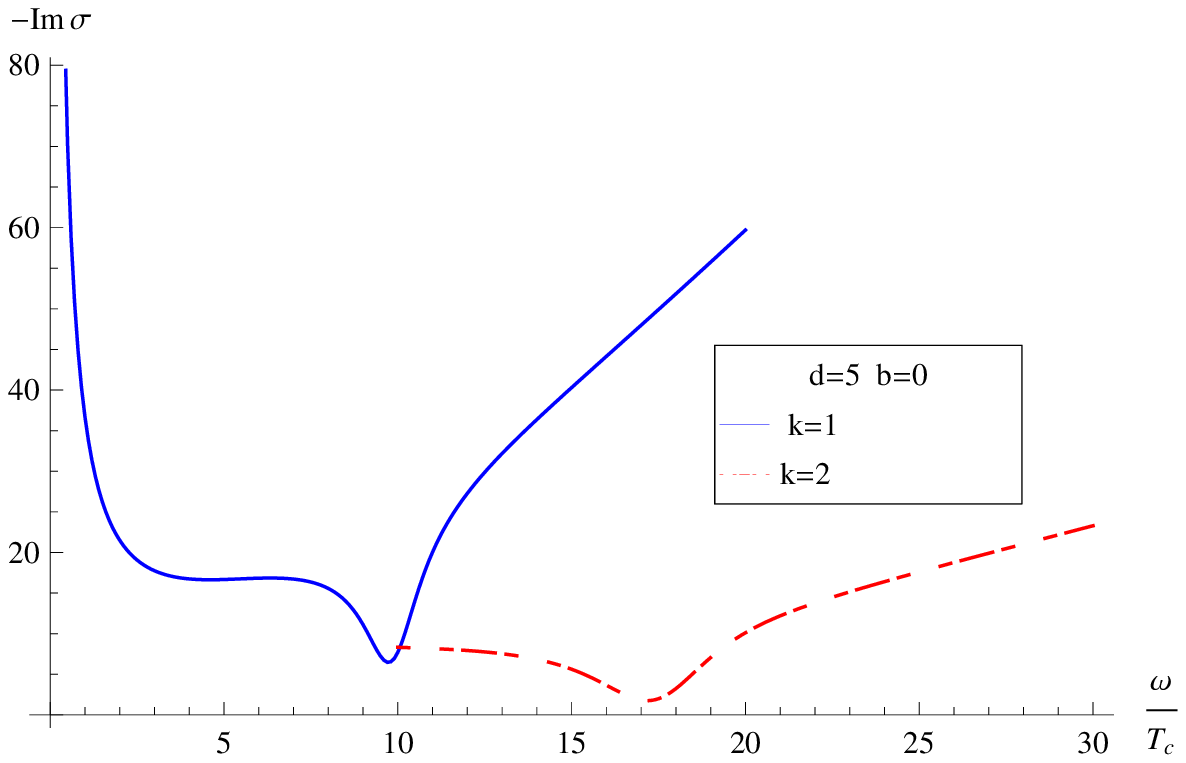}
\caption{Real(left) and imaginary(right) parts of $\langle\mathcal{O}_{1}\rangle$ conductivity for zero black hole charge in five dimensions and varying $k$.}
\label{fixd5_O1}
\end{figure}

\begin{figure}[htb!]
\centering
\includegraphics[scale=0.6]{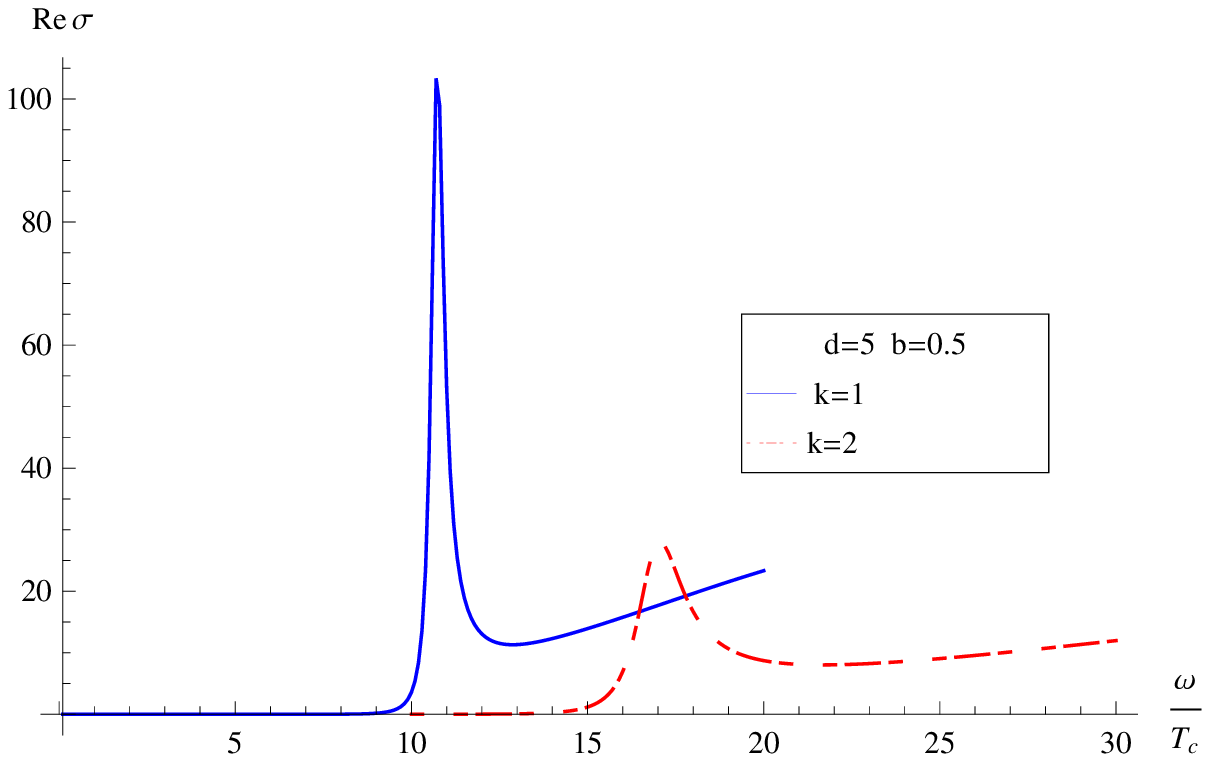}
\includegraphics[scale=0.6]{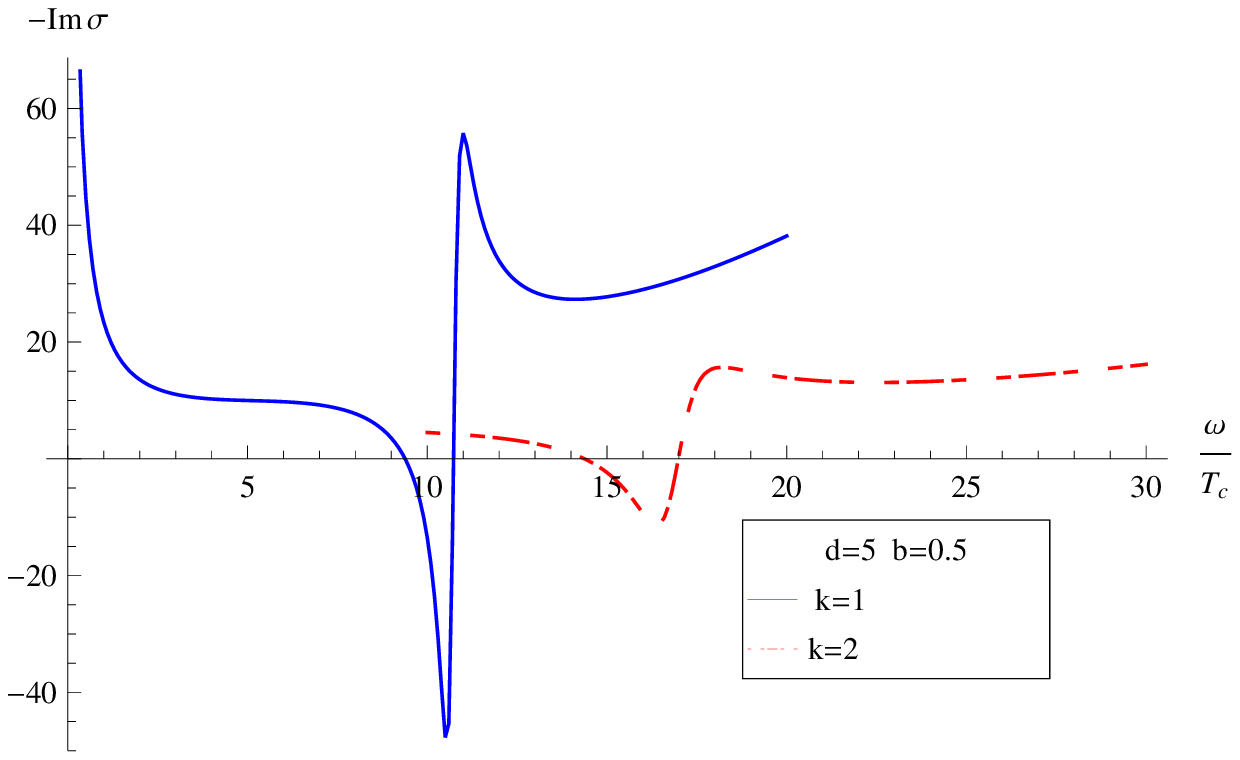}
\caption{Real(left) and imaginary(right) parts of {\bf{$\langle\mathcal{O}_{2}\rangle$}} conductivity for charged black hole in five dimensions and varying $k$.}
\label{fixd5_charge_O2}
\end{figure}

Having these two quantities, we compute the conductivity $\sigma(\omega)$ through the Ohm's law
\begin{equation}
\sigma(\omega)=-\frac{i\langle J_{\mu}\rangle}{\omega A_{x}}.
\end{equation}
We present in Fig.(\ref{fixd5_O1})-Fig.(\ref{fixd5_charge_O2}) the real end imaginary part of conductivity $\sigma(\omega)$ of $\langle\mathcal{O}_{1}\rangle$ and $\langle\mathcal{O}_{2}\rangle$ for five dimensional black hole in charged and uncharged cases. The conductivity phenomena is qualitatively very similar in both cases $k=1$ and $k=2$ for the two operators, but the $k=2$ corrections to the curvature seems to lower the conductivity comparing to the $k=1$ case.

\begin{figure}[htb!]
\centering
\includegraphics[scale=0.6]{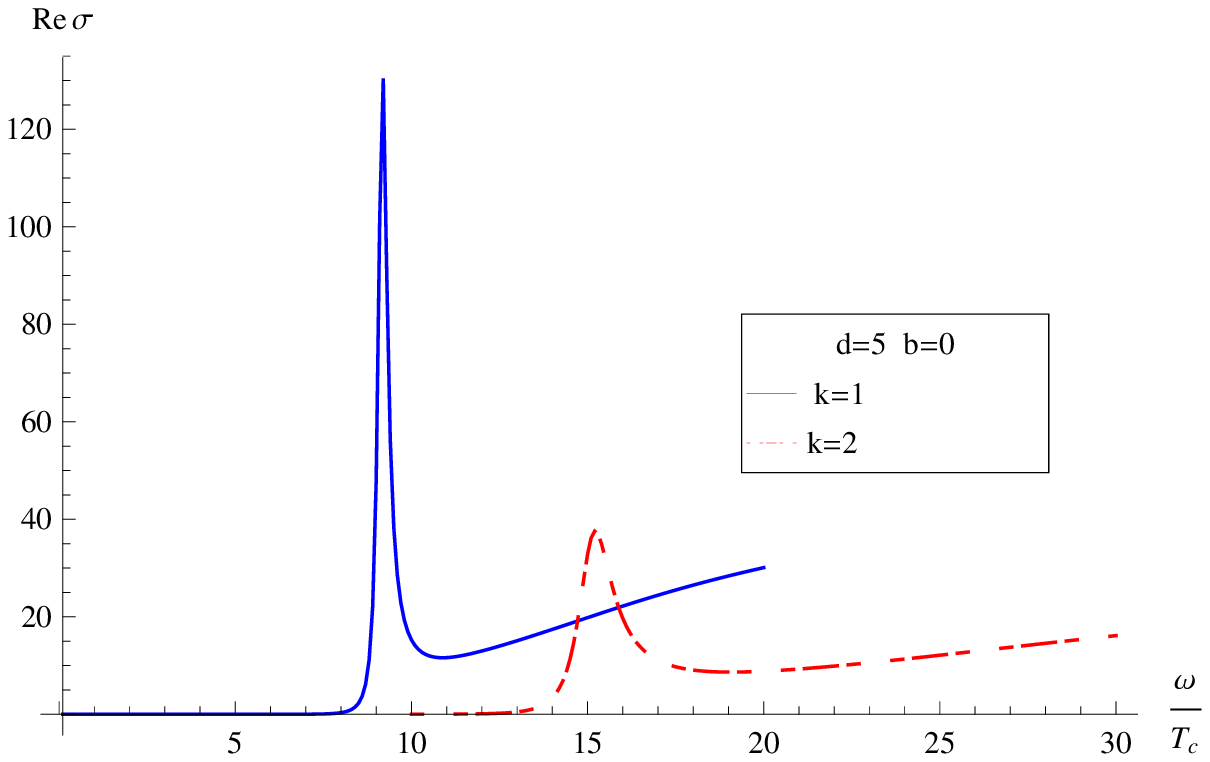}
\includegraphics[scale=0.6]{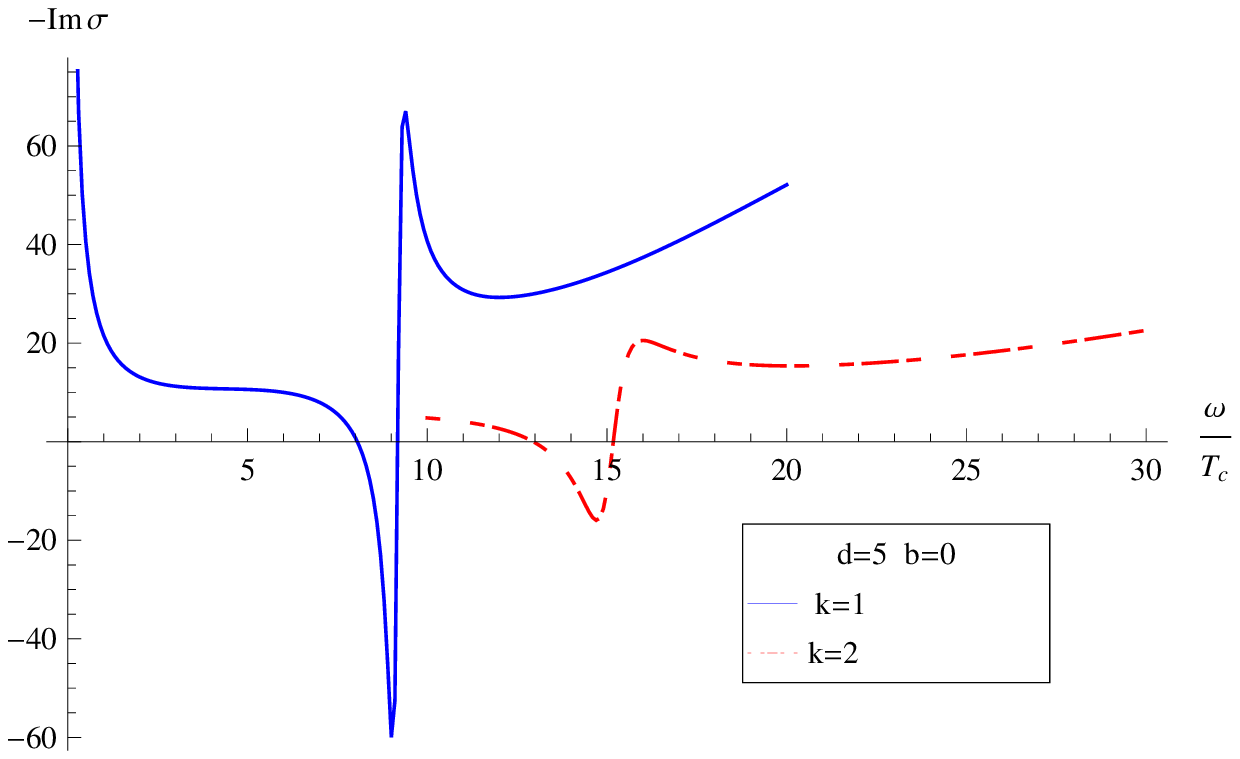}
\caption{Real(left) and imaginary(right) parts of {\bf{$\langle\mathcal{O}_{2}\rangle$}} conductivity for zero black hole charge in five dimensions and varying $k$.}
\label{fixd5_O2}
\end{figure}

\begin{figure}[htb!]
\centering
\includegraphics[scale=0.6]{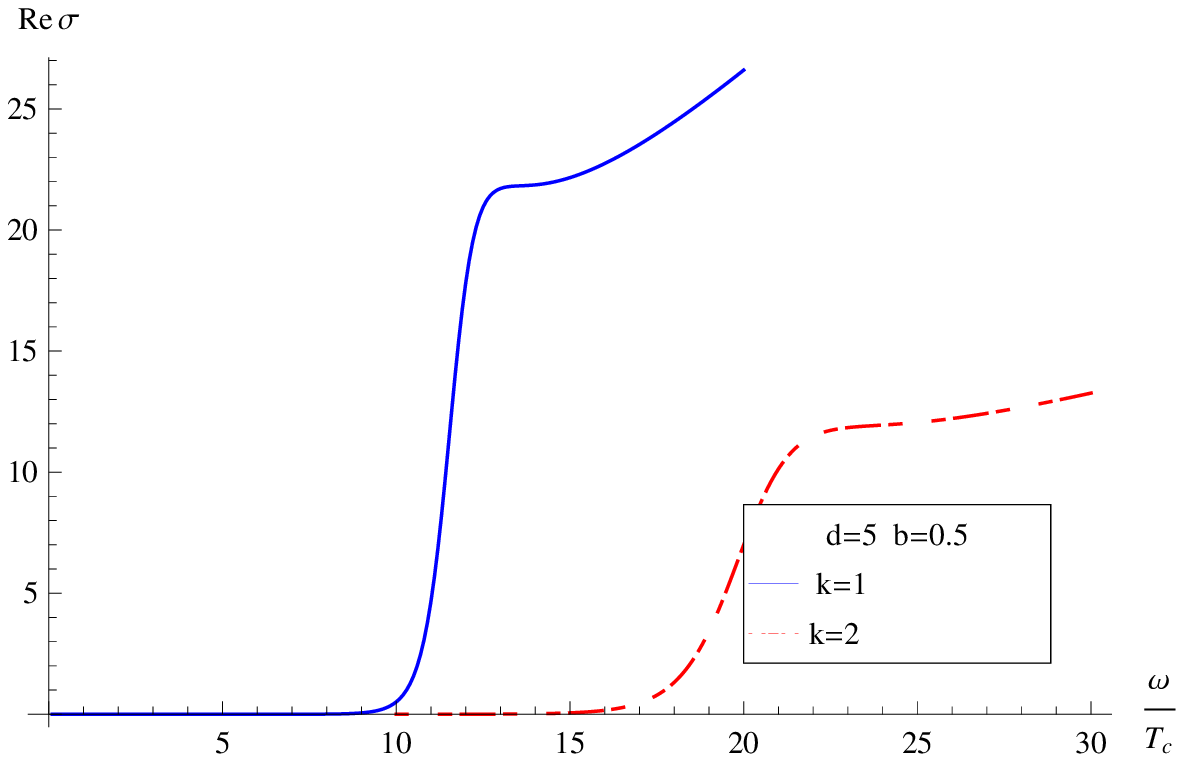}
\includegraphics[scale=0.6]{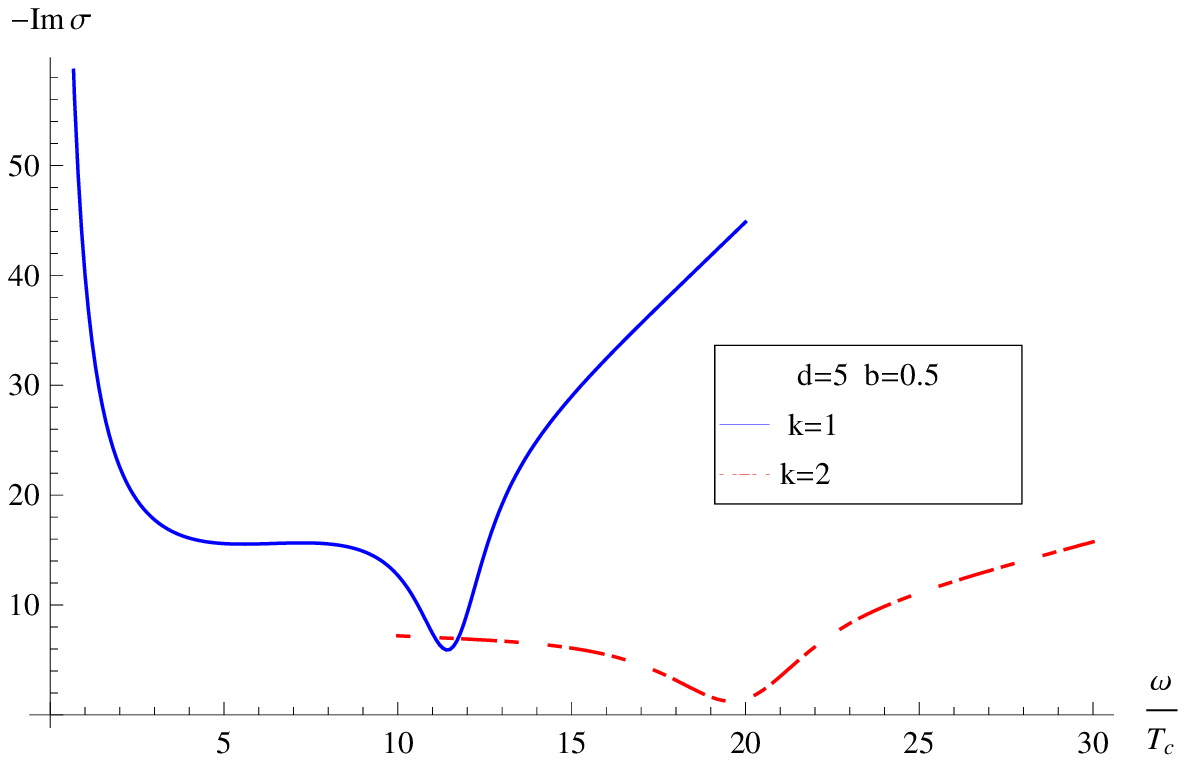}
\caption{Real(left) and imaginary(right) parts of {\bf{$\langle\mathcal{O}_{1}\rangle$}} conductivity for charged black hole in five dimensions and varying $k$.}
\label{fixd5_charge_O1}
\end{figure}

\section{$R$-current correlators and hydrodynamical quasinormal modes }\label{sec:r_current}

In this section we are going to apply the AdS/CFT correspondence
\cite{witten} \cite{Kovtun:2005ev} in order to compute the real time
$R-$current correlators, which can be expressed in terms of the
boundary value of the gauge invariant quantities such as the
electric field at the spatial infinity. As one knows, the
electromagnetic fluctuations, in the AdS/CFT context, give rise to
the correlators associated to the $R-$symmetry at the boundary field
theory.

Following the procedure outlined in \cite{Kovtun:2005ev}
\cite{Miranda:2008vb}, the imposition of Dirichlet boundary
conditions on the gauge invariant variables lead to the poles of the
field theory correlation functions and, according to
\cite{Nunez:2003eq},  the quasinormal frequency spectra of the
asymptotic AdS black hole considered. Moreover, a consequence of
applying the approach \cite{Nunez:2003eq} is that the
electromagnetic  quasinormal spectra presents a set of modes which
behaves like a diffusion wave in the long wavelength and low
frequency limit, such a limit is called hydrodynamic limit of
perturbations. The two main results of the section is the explicit
form of correlators in the field theory defined at the boundary of
Lovelock black holes and the frequency of diffusion quasinormal
modes for dimension $d\geq 4$.

\subsection{Correlators due to Electromagnetic Field}


We are going to consider as our bulk geometry the uncharged
$d-$dimensional planar Lovelock black hole, represented by the
following line element
\begin{equation}\label{metric_u}
ds^{2}=\frac{r_{0}^{2}}{l^{2}u^{2/(d-3)}}\left[-g(u)dt^{2}+\sum_{i}^{d-3}dx_{i}^{2}\right]+\frac{r_{0}^{2}}{u^{2/(d-3)}}d\phi^{2}+\frac{l^{2}}{(d-3)^{2}u^{2}g(u)}du^{2}.
\end{equation}
The function $g(u)$ is the horizon function given by
\[
g(u)=1-u^{\gamma}, \hspace{0.3cm} \gamma=\frac{(d-1)}{k(d-3)}.
\]

The event horizon is located at $r=r_{0}$ or $u=1$, the radial
coordinate $r \in [r_{0},+\infty]$ is mapped to $u \in [1,0]$
through $u=r_{0}/r$. In order to have a black hole with planar
topology, at least one of the extra dimensions has to be compact
\cite{arostroncosozanelli}, so in the above metric $\phi \in
[0,2\pi]$, and the remaining directions have the domain
$x_{i}\in[-\infty,+\infty]$, where $i=1\cdots d-3$.

In AdS/CFT context, the electromagnetic field evolving in the region
near the AdS boundary  couples to the holographic field theory
current-current  two-point correlation function. Therefore, we have
to compute the perturbations on the electromagnetic potential
$A_{\mu}$, whose equations governing its dynamics are the Maxwell
equations,
\begin{equation}\label{maxwell}
\partial_{\mu}\left(\sqrt{-g}F^{\mu\nu}\right)=0,
\end{equation}
where $F_{\mu\nu}=\partial_{\mu}A_{\nu}-\partial_{\nu}A_{\mu}$, and
the metric components which enter in Maxwell equation are those
given by (\ref{metric_u}). Using the isometries of black hole
spacetime, we can decompose the gauge field $A_{\mu}$ in Fourier as
following
\begin{equation}
A_{\mu}(t,x_{i},\phi,u)=\frac{1}{(2\pi)^{d-1}}\int
(dw)(dm)(dq_{i})^{d-3}e^{-i\omega t +im\phi
+iq_{i}x^{i}}A_{\mu}(\omega,m,q_{i},u).
\end{equation}

It is possible, without loss of generality, choose a $(d-1)-$
dimensional wave vector $\vec{p}=(-\omega,Q_{a})$(with $a=1,\cdots,d-2$), where
$Q_{a}=(m,q_{i})=(0,q,\vec{0})$, so, as initial configuration,
have the gauge field $A_{\mu}$ propagating in one of the planar
directions $x_{i}=(x,\vec{0})$. Such choice allows us to consider the
perturbations on the gauge field as two orthogonal sets
\cite{chandra} \cite{Miranda:2008vb}, the odd perturbations
$A_{\phi}$ and the even perturbations $A_{t}, A_{x}, A_{u}$.

Our gauge choice is the radial gauge where $A_{r}=A_{u}=0$, and the
fundamental gauge invariant variables for the two classes of
perturbations are the transverse component of electric field
$E_{\phi}$ for the odd perturbations and the component $E_{x}$ for
the even perturbations. The equations governing the dynamics are
obtained from the Maxwell equations (\ref{maxwell}) written on the
spacetime (\ref{metric_u}):
\begin{eqnarray}\label{odd_electric}
E_{\phi}''+\frac{g(u)'}{g(u)}E_{\phi}'+\frac{\mathfrak{w}^{2}-\mathfrak{q}^{2}g(u)}{(d-3)^{2}g(u)^{2}u^{2\frac{d-4}{d-3}}}E_{\phi}
&=&0\quad ,\\
\label{even_electric}
E_{x}''+\frac{g(u)'\mathfrak{w}^{2}}{g(u)\left[\mathfrak{w}^{2}-\mathfrak{q}^{2}g(u)\right]}E_{x}'
++\frac{\mathfrak{w}^{2}-\mathfrak{q}^{2}g(u)}{(d-3)^{2}g(u)^{2}u^{2\frac{d-4}{d-3}}}E_{x}&=&0\quad ,
\end{eqnarray}
where the primes refers to derivatives with respect to $u$
direction. For convenience, we have normalized the quantities
$\mathfrak{w}$ and $\mathfrak{q}$ in terms of black hole Hawking
temperature
\[
T=\frac{(d-1)}{4\pi l^{2}k}r_{0}\quad ,
\]
namely,
\[
\mathfrak{w}=\frac{(d-1)}{4\pi k}\frac{\omega}{T}\quad , \hspace{0.3cm}
\mathfrak{q}=\frac{(d-1)}{4\pi k}\frac{q}{T}\quad .
\]

Following the AdS/CFT recipe \cite{sonstarinets}, the current-current
two point correlators are given by the field $E_{\mu}$
($\mu=\phi,x$) near the AdS boundary, which in our case, is obtained
through the solution of equations (\ref{odd_electric}) and
(\ref{even_electric}) at $u\approx 0$:
\begin{eqnarray}\label{bdry_ephi}
E_{\phi}&=&a_{\phi}(\mathfrak{w},\mathfrak{q})+b_{\phi}(\mathfrak{w},\mathfrak{q})u\quad ,\\
\label{bdry_ex}
E_{x}&=&a_{x}(\mathfrak{w},\mathfrak{q})+b_{x}(\mathfrak{w},\mathfrak{q})u\quad ,
\end{eqnarray}
furthermore, close to the event horizon $E_{\mu}=g(u)^{\pm\frac{i k
}{d-1}\mathfrak{w}}$, where the positive exponent corresponds to
outgoing waves and the negative exponent to ingoing waves at the
event horizon, also the choice of sign means the electric field at
AdS boundary is taken as classical source of retarded (negative) or
advanced (positive) current-current two point correlators. The
ingoing waves at the event horizon are physically motivated boundary
conditions for a classical black hole, thus, we are going to adopt
the negative exponent for the electrical field $E_{\mu}$ meaning
that we are considering the retarded correlators of the holographic
field theory.

The next step is to consider the electromagnetic action at the AdS
boundary ($u\approx 0$) with the results (\ref{bdry_ephi}) and
(\ref{bdry_ex}) we have
\begin{equation}\label{bdry_action}
S=\frac{(d-3)r_{0}^{(d-3)}}{2\eta^{2} l^{d-3}}\int \frac{d\omega
dq}{(2\pi)^{2}}\left[\frac{g(u)}{\mathfrak{q}^{2}g(u)-\mathfrak{w}^{2}}E_{x}(u,
-\vec{p})E_{x}'(u,\vec{p})-\frac{g(u)}{\mathfrak{w}^{2}}E_{\phi}(u,
-\vec{p})E_{\phi}'(u,\vec{p})\right].
\end{equation}
where $\eta^{2}$ is the normalization of the action, from
\cite{Freedman:1998tz} one finds
\[
\frac{1}{\eta^{2}}=\frac{(d-1)\Gamma[\frac{d}{2}]}{2^{(d-1)}\pi^{\frac{d}{2}}\Gamma[d]}(N_{c}^{2}-1)\quad ,
\]
with $N_{c}$ representing the number of $D-$ branes. Also, we can
rewrite the electric field at the AdS boundary in terms of the gauge
field in the same region $A^{0}_{\mu}(\vec{p})=A_{\mu}(u\rightarrow
0, \vec{p})$ and applying the Lorentzian prescription \cite{sonstarinets},
\begin{equation}
C_{\mu\nu}(\omega,\vec{p})=\frac{2\delta^{2}S}{\delta
A^{0}_{\mu}(\vec{p})\delta A_{\nu}^{0}(-\vec{p})}\quad ,
\end{equation}
 one finds the current-current correlators
\begin{eqnarray}
C_{tt}(\omega,q)&=&\frac{(d-3)r_{0}^{d-3}}{\eta^{2} l^{d-3}}\frac{b_{x}(\mathfrak{w},\mathfrak{q})}{a_{x}(\mathfrak{w},\mathfrak{q})}\frac{\mathfrak{q}^{2}}{\mathfrak{w}^{2}-\mathfrak{q}^{2}},\\
C_{xx}(\omega,q)&=&\frac{(d-3)r_{0}^{d-3}}{\eta^{2} l^{d-3}}\frac{b_{x}(\mathfrak{w},\mathfrak{q})}{a_{x}(\mathfrak{w},\mathfrak{q})}\frac{\mathfrak{w}^{2}}{\mathfrak{w}^{2}-\mathfrak{q}^{2}},\\
C_{\phi\phi}(\omega,q)&=&\frac{(d-3)r_{0}^{d-3}}{\eta^{2} l^{d-3}}\frac{b_{\phi}(\mathfrak{w},\mathfrak{q})}{a_{\phi}(\mathfrak{w},\mathfrak{q})},\\
C_{tx}(\omega,q)&=&\frac{(d-3)r_{0}^{d-3}}{\eta^{2}
l^{d-3}}\frac{b_{x}(\mathfrak{w},\mathfrak{q})}{a_{x}(\mathfrak{w},\mathfrak{q})}\frac{\mathfrak{w}\mathfrak{q}}{\mathfrak{w}^{2}-\mathfrak{q}^{2}}.
\end{eqnarray}
Using the components of $C_{\mu\nu}$ it is possible to express the
transversal  $\Pi^{T}(\omega,q)$ and longitudinal
$\Pi^{L}(\omega,q)$ self-energies of the $(d-1)$ holographic thermal
field theory
\begin{eqnarray}
\Pi^{T}(\omega,q)=\frac{(d-3)r_{0}^{d-3}}{\eta^{2} l^{d-3}}\frac{b_{\phi}(\mathfrak{w},\mathfrak{q})}{a_{\phi}(\mathfrak{w},\mathfrak{q})},\\
\Pi^{L}(\omega,q)=\frac{(d-3)r_{0}^{d-3}}{\eta^{2}
l^{d-3}}\frac{b_{x}(\mathfrak{w},\mathfrak{q})}{a_{x}(\mathfrak{w},\mathfrak{q})}.
\end{eqnarray}
Therefore, the electromagnetic correlation functions are fully determined
by the relations
$b_{\phi}(\mathfrak{w},\mathfrak{q})/a_{\phi}(\mathfrak{w},\mathfrak{q})$
and
$b_{x}(\mathfrak{w},\mathfrak{q})/a_{x}(\mathfrak{w},\mathfrak{q})$
and the poles of the correlators are the same as the zeros of
$a_{\phi}(\mathfrak{w},\mathfrak{q})$ and
$a_{x}(\mathfrak{w},\mathfrak{q})$ \cite{Nunez:2003eq}.  To find the
poles, we impose Dirichlet boundary conditions on the electric field
at AdS boundary and ingoing wave conditions at the black hole event
horizon.

\subsection{Diffusion Quasinormal modes }

To determine the self-energies found in the preceding computation,
we have to solve the differential equations for $E_{x}$ and
$E_{\phi}$. Analytical solutions are unknown, unless in the so-called
{\it{hydrodynamical limit}} of the perturbations. Such a limit is
achieved by considering a set of perturbations with small
frequencies and small wave numbers,
\[
\mathfrak{w}\ll 1,\hspace{0.3cm} \mathfrak{q}\ll 1.
\]

From the point of view of the thermal field theory, at least one of
the electromagnetic quasinormal frequencies has to behave as a
diffusion mode in the hydrodynamical limit. So, if we impose
Dirichlet and ingoing-wave boundary conditions to the differential
equations (\ref{odd_electric}) and (\ref{even_electric}), we found
that there is not a transversal diffusion mode, namely, does not
exist a value of $\omega$ that is compatible with $E_{\phi}=0$ at
AdS boundary. Such a result is independent on the dimensionality of
the bulk and the flavor of the Lovelock theory, in other words,
independent on $d$ and $k$. However, we found that for the
longitudinal mode, there is a hydrodynamical mode given by
\begin{equation}\label{hydro_mode}
\mathfrak{w}=-i\frac{\mathfrak{q}^{2}}{(d-3)}\Rightarrow
\omega=-i\frac{(d-1)}{4\pi(d-3)k T}q^{2},
\end{equation}
whose diffusion coefficient can be read off
\begin{equation}\label{diffusion_mode}
D=\frac{(d-1)}{4\pi(d-3)k T}.
\end{equation}
This is the main result of the section. We found that the diffusion
coefficient depends crucially on the flavor of Lovelock gravity. As
we increase the corrections to the curvature in Lovelock Lagrangian
the diffusion coefficient tend to zero, so the charge diffusion in
longitudinal direction in thermal field theory is diminished in
gravity duals with corrections to the curvature.

\section{Concluding remarks}\label{sec:concluding}

In this work we have studied the effects of higher order corrections to the gravity upon the scalar and hydrodynamical quasinormal modes spectrum, the condensation of holographic operators and their conductivity.

Regarding to the scalar quasinormal modes, we found that the
corrections to the curvature diminish the quasinormal  modes
oscillating phase, it is similar to the dynamics of a perturbation in
a very dense material medium. We see from Fig.\ref{variousdim52} the
case where the real part of the frequencies are zero, so these modes
are purely damped. Moreover, we found in the hydrodynamical limit a
purely damped diffusive quasinormal mode $\omega=-i(d-1)q^{2}/4\pi(d-3)k
T$, which depends strongly on the $k$ parameter.

We obtained explicitly the phase transition giving the condensation of
operators $\langle\mathcal{O}_{1}\rangle$ and
$\langle\mathcal{O}_{2}\rangle$. The influence of curvature
corrections of the Lovelock gravity is to increase the value of the
condensate, in both charged and uncharged cases. Also, we compute the
conductivity, where we found that the considered gravity bulk diminish
the real part and imaginary part of $\sigma(\omega)$ as we add more
corrections to curvature.

As an extension of this work, it would be interesting to consider
charged fermions fields evolving on the gravity bulk given by the
family of Lovelock black holes in order to investigate if purely damped
quasinormal frequencies  are allowed in this case. Another
problem which will be address in a future work is the question of
gravitational stability of Lovelock black holes and the computation of
holographic stress-energy tensor of field theory on the spacetime AdS boundary.

\appendix
\section{Equations of motion for $(d=5,k=2)$, $(d=6,k=1)$ and $(d=6,k=2)$}
The general equations of motion of the scalar and gauge fields are
\begin{eqnarray}
&&\psi_{5,2}''+\frac{\left(\left(b^4+b^2+1\right) z^2-3 \sqrt{1-b^2
\left(b^2+1\right) \left(z^2-1\right)}\right) }{z \sqrt{1-b^2
\left(b^2+1\right) \left(z^2-1\right)}+\left(b^4+b^2\right)
z^5-\left(b^4+b^2+1\right)z^3}\psi_{5,2}'\nb\\
&&~~~~~~~~~~~~+\frac{ \left(m^2 \left(z^2 \sqrt{1-b^2
\left(b^2+1\right) \left(z^2-1\right)}-1\right)+z^2
\varphi_{5,2}^2\right)}{z^2 \left(z^2 \sqrt{1-b^2 \left(b^2+1\right)
\left(z^2-1\right)}-1\right)^2}\psi_{5,2}=0,\nb\\
&&\varphi_{5,2}''-\frac{\varphi_{5,2}'}{z}+\frac{2
\varphi_{5,2}\psi_{5,2}^2}{z^2 \left(z^2 \sqrt{1-b^2
\left(b^2+1\right) \left(z^2-1\right)}-1\right)}=0,\nb\\
&&A_{x,5,2}''+\left[\frac{\left(\frac{\left(b^4+b^2\right)
z}{\sqrt{1-b^2 \left(b^2+1\right)
\left(z^2-1\right)}}-\frac{2}{z^3}\right)}{\frac{1}{z^2}-\sqrt{1-b^2
\left(b^2+1\right)
\left(z^2-1\right)}}+\frac{1}{z}\right]A_{x,5,2}'\nb\\
&&~~~~~~~~~~~~~~~+\frac{2 \left(z^2 \sqrt{1-b^2 \left(b^2+1\right)
\left(z^2-1\right)}-1\right) \psi_{5,2}^2+\omega ^2 z^2}{z^2
\left(z^2 \sqrt{1-b^2 \left(b^2+1\right)
\left(z^2-1\right)}-1\right)^2}A_{x,5,2}=0.\nb\\
&&\psi_{6,1}''+\frac{-2 \left(b^2+b+1\right)-3 (b+1)
\left(b^2+1\right) \left(b^4+1\right) z^5+6 b^3
\left(b^4+b^3+b^2+b+1\right) z^8}{z(b^2-(b+1) \left(b^2+1\right)
\left(b^4+1\right) z^5+b^3
\left(b^4+b^3+b^2+b+1\right) z^8+b+1)}-\frac{2}{z}\psi_{6,1}'\nb\\
&&~~~~~~~~~~~~~+\frac{\varphi_{6,1}^2-m^2 \left[\frac{1}{z^2}-z^6
\left(\frac{b^8-1}{z^3(b^3-1)}+\frac{b^3-b^8}{b^3-1}\right)\right]}{z^4
\left[\frac{1}{z^2}-z^6
\left(\frac{b^8-1}{z^3(b^3-1)}+\frac{b^3-b^8}{b^3-1}\right)\right]^2}\psi_{6,1}=0,\nb\\
&&\varphi_{6,1}''-\frac{2}{z}\varphi_{6,1}'-\frac{2
(b^3-1)\varphi_{6,1}\psi_{6,1}^2}{(b^3-1)z^2-(b^8-1)z^7+(b^3-b^8)z^{10}}=0,\nb\\
&&A_{x,6,1}''+\frac{\frac{3 \left(1-b^8\right) z^2}{b^3-1}+\frac{6
b^3 \left(b^4+b^3+b^2+b+1\right)
z^5}{b^2+b+1}-\frac{2}{z^3}}{\frac{1}{z^2}-z^6
\left(\frac{b^8-1}{z^3(b^3-1)}+\frac{b^3-b^8}{b^3-1}\right)}A_{x,6,1}'\frac{\omega
^2-2 \left[\frac{1}{z^2}-z^6
\left(\frac{b^8-1}{z^3(b^3-1)}+\frac{b^3-b^8}{b^3-1}\right)\right]\psi
(z)^2}{z^4\left[\frac{1}{z^2}-z^6
\left(\frac{b^8-1}{z^3(b^3-1)}+\frac{b^3-b^8}{b^3-1}\right)\right]^2}A_{x,6,1}=0.\nb\\
&&\psi_{6,2}''-\left(\frac{\frac{b^8-4 \left(b^5-1\right)b^3
z^3-1}{2 \sqrt{b^3-1} \sqrt{z \left(b^8-\left(b^5-1\right) b^3
z^3-1\right)}}+\frac{2}{z^3}}{\frac{1}{z^2}-\sqrt{\frac{z
\left(b^8-\left(b^5-1\right) b^3
z^3-1\right)}{b^3-1}}}+\frac{2}{z}\right)\psi_{6,2}'\frac{m^2 z^2
\sqrt{\frac{z \left(b^8-\left(b^5-1\right) b^3
z^3-1\right)}{b^3-1}}-m^2+z^2 \varphi_{6,2}^2}{z^2
\left(z^2\sqrt{\frac{z \left(b^8-\left(b^5-1\right) b^3
z^3-1\right)}{b^3-1}}-1\right)^2}\psi_{6,2}=0,\nb\\
&&\varphi_{6,2}''-\frac{2 }{z}\varphi_{6,2}'+\frac{2 \varphi_{6,2}
\psi_{6,2}^2}{z^4 \sqrt{\frac{z \left(b^8-\left(b^5-1\right) b^3
z^3-1\right)}{b^3-1}}-z^2}=0,\nb\\
&&A_{x,6,2}''+\frac{-\frac{b^8-4 \left(b^5-1\right) b^3 z^3-1}{2
\left(b^3-1\right) \sqrt{\frac{z\left(b^8-\left(b^5-1\right) b^3
z^3-1\right)}{b^3-1}}}-\frac{2}{z^3}}{\frac{1}{z^2}-\sqrt{\frac{z
\left(b^8-\left(b^5-1\right) b^3
z^3-1\right)}{b^3-1}}}A_{x,6,2}'\frac{2 z^2 \sqrt{\frac{z
\left(b^8-\left(b^5-1\right) b^3 z^3-1\right)}{b^3-1}}
\psi_{6,2}^2+\omega ^2 z^2-2 \psi_{6,2}^2}{z^2\left(z^2
\sqrt{\frac{z \left(b^8-\left(b^5-1\right)
b^3z^3-1\right)}{b^3-1}}-1\right)^2}A_{x,6,2}=0
\end{eqnarray}

We get the results for different values of $b$ (representing the
charge) $k$ and the dimension $d$.

\section*{\bf Acknowledgements}

This work was supported by FAPESP No. 2012/08934-0 and CNPq, Brazil.


\onecolumngrid

\end{document}